\documentclass{aa}
\usepackage[dvipsnames]{xcolor}
\usepackage{natbib}
\usepackage{graphicx}
\usepackage{stfloats}
\usepackage{placeins}
\usepackage[flushleft]{threeparttable}
\usepackage{tikz}

\usepackage{pdflscape}
\usepackage{rotating}
\usepackage{txfonts}
\usepackage[normalem]{ulem}
\usepackage[breaklinks,colorlinks,urlcolor=blue,citecolor=blue,linkcolor=blue]{hyperref}
\begin{document}

   \title{Impact of accretion variability on the CO emission \\ of a disk around a very-low-mass star with JWST}

   \author{C. Bergez-Casalou\inst{1}, B. Tabone\inst{1}, Y. Aoyama \inst{2}, G. Herczeg\inst{3,4}, A. Caratti o Garatti\inst{5}, E. Habart \inst{1}, E. F. van Dishoeck\inst{6,7}, I. Chemerynska \inst{1} , T. Henning\inst{8},    
   S. L. Grant\inst{9}, M. Guedel\inst{10,11}, A. M. Arabhavi\inst{12}, J. Kanwar\inst{13}, G. Olofsson\inst{14}, G. Perotti\inst{8,15}, L. Tychoniec\inst{6}, L. B. F. M. Waters\inst{16,17}} 

   \institute{\inst{1}Université Paris-Saclay, CNRS, Institut d’Astrophysique Spatiale, 91405 Orsay, France,  email: camille.bergez@obspm.fr\\
   \inst{2}School of Physics and Astronomy, Sun Yat-sen University, Zhuhai 519082, People's Republic of China \\
   \inst{3}Kavli Institute for Astronomy and Astrophysics, Peking University, Yiheyuan Lu 5, Haidian Qu, 100871 Beijing, People’s Republic of China\\
   \inst{4}Department of Astronomy, Peking University, Yiheyuan 5, Haidian Qu, 100871 Beijing, People’s Republic of China\\
   \inst{5}INAF – Osservatorio Astronomico di Capodimonte, Salita Moiariello 16, 80131 Napoli, Italy \\  
   \inst{6}Leiden Observatory, Leiden University, PO Box 9513, 2300 RA Leiden, The Netherlands \\
   \inst{7}Max-Planck-Institut für Extraterrestrische Physik, Giessenbachstrasse1, 85748 Garching, Germany \\
   \inst{8}Max-Planck-Institut für Astronomie, Königstuhl 17, 69117 Heidelberg, Germany \\
   \inst{9}Earth and Planets Laboratory, Carnegie Institution for Science, 5241 Broad Branch Road, NW, Washington, DC 20015, USA\\
   \inst{10}Dept. of Astrophysics, University of Vienna, Türkenschanzstr. 17, A-1180 Vienna, Austria \\
   \inst{11}ETH Zürich, Institute for Particle Physics and Astrophysics, Wolfgang-Pauli-Str. 27, 8093 Zürich, Switzerland \\
   \inst{12}Kapteyn Astronomical Institute, Rijksuniversiteit Groningen, Post-bus 800, 9700AV Groningen, The Netherlands \\
   \inst{13}Department of Astronomy, University of Michigan, 1085 S. University Ave, Ann Arbor, MI 48109, USA \\
   \inst{14}Department of Astronomy, Stockholm University, AlbaNova University Center, 10691, Stockholm, Sweden  \\
   \inst{15}Niels Bohr Institute, University of Copenhagen, NBB BA2, Jagtvej 155A, 2200 Copenhagen, Denmark \\
   \inst{16}HFML-FELIXToernooiveld 7, Nijmegen, 6525 ED, the Netherlands\\
   \inst{17}Department of Astrophysics, IMAPP, Radboud University, Nijmegen, The Netherlands\\}


 
 \abstract
  {\textit{Context:} Very-low-mass stars ($M_\star<0.2M_\odot$) are known to host a larger fraction of terrestrial planets compared to higher mass stars. Studying the structure of their protoplanetary disk is therefore crucial to understand the formation of Earth-like planets.
  
  \textit{Aims:} In this paper, we aim to characterize the innermost regions of the disk around a known very-low-mass star, 2MASS J16053215-1933159 (J1605 hereafter), by analyzing new JWST/NIRSpec observations taken in 2024, 2 years after a first observation with JWST/MIRI in 2022. We investigate the atomic and molecular emission in both spectra to establish a global view of the system and study its variability.
  
  \textit{Methods:} After identifying both the fundamental band of CO and HI recombination emission lines from both spectra corrected from the contribution of the stellar photosphere, we use the HI lines to derive the stellar accretion rate at each epoch. The CO emission is fitted with a slab model to derive the gas characteristics. 
  
  \textit{Results:} We find that between the two epochs (2022 \& 2024), the CO flux is reduced by a factor $\sim 3$, the flux of the only HI line seen by both NIRSpec and MIRI (the HI 10-6 line) is reduced by a factor $\sim13$, and the total continuum density flux is reduced by $\sim 15 \%$. The accretion rate derived from individual HI lines is consistent with a decrease of a factor $\sim13$ in 2 years. After correcting the spectrum from the CO absorption present in J1605 stellar photosphere, we find that the decrease of the disk's CO emission is consistent with a CO gas of same column density and temperature at each epoch, but originating from a smaller emitting area when the accretion rate is low (NIRSpec's epoch). While warm hydrocarbons ($\simeq 500~$K) are detected with MIRI-MRS (2022) with extremely high column densities of $\rm C_2H_2$, no hydrocarbons features are seen with NIRSpec (2024). We propose that non-LTE effects quench the near-IR emission of hydrocarbons in the warm reservoir. Taking the newly detected hot CO ($\sim$ 800~K) as a reference and assuming LTE, we constrain the column density of $\rm C_2H_2$, HCN and $\rm CH_4$ to be less than 30\% of CO in the hot reservoir. This value suggests that the C/O in the inner disk is unlikely to be much larger than unity.
    
  \textit{Conclusion:} The variation of the accretion luminosity of the very-low-mass star J1605 correlates with the variation of its fundamental CO luminosity in a similar way as in T-Tauri stars, supporting the idea that very-low-mass stars could be seen as scaled-down versions of T-Tauri stars. The detection of the emission of the fundamental band of CO in disks around cold objects (very-low-mass stars, brown dwarfs and giant planets) is challenging because of the presence of CO absorption in their photosphere. Future observations (e.g., with ELT/METIS) looking for such emission will require robust ways to constrain the photosphere emission and follow the time variability of the sources. 
  
  }
  


   \authorrunning{C. Bergez-Casalou et al}
   \titlerunning{Accretion variability in a VLMS}

   \maketitle
%

\section{Introduction}

Very-low-mass stars (M-dwarfs, $M_\star<0.2M_\odot$) are the most common type of stars in our galaxy \citep[see review by][]{Henry2024}. They are known to host a large fraction of terrestrial planets \citep[e.g.,][]{Ment2023}. The architecture, final mass and composition of these resulting planetary systems will be inherited from the protoplanetary disk where they formed. Observations of protoplanetary disks are therefore crucial to understand the characteristics of the birth environment of these planets. In particular, infrared observations give us access to the emission of the inner part of the disks, where the majority of the detected planets are expected to form. 

Recent JWST/MIRI observations have shown that the emission of disks around very-low-mass stars is almost always dominated by hydrocarbons compared to the disks around higher mass T-Tauri stars (\citealt{Pascucci2013,Tabone2023,Arabhavi2024,Grant2025,Arabhavi2025,Arabhavi2025b,Long2025}, with the exception of the variable brown dwarf system J0438, \citealt{Perotti2026} and of Sz114, \citealt{Xie2023}). This peculiarity could be explained by their small stellar luminosity and disk's size and mass: as their icelines are located much closer to the central star \citep[e.g.,][]{Greenwood2017} and dynamical timescales are shorter \citep[e.g.,][]{Pinilla2013}, the transport of material towards the inner disk has a different impact on the disk's composition than for disks around higher mass T-Tauri stars \citep{Mah2023}. Disks around very-low-mass stars are therefore unique, making their study important for the formation of terrestrial planets. 

The atomic and molecular emission of the inner disk is tightly linked to the stellar accretion process. During magnetospheric accretion \citep[e.g.,][]{Pringle1972,Hartmann1994,Bouvier2007,Romanova2015}, the accreted material is shocked at the stellar surface after being transported from the inner disk via funnel flows, producing an accretion luminosity $L_{\rm acc}$ able to heat the inner disk via a UV excess. $L_{\rm acc}$ can be derived directly from the observation of this UV continuum excess \citep[e.g.,][]{Gullbring1998,Herczeg2008} or from the hydrogen recombination emission lines produced by the accretion shock both in the optical \citep[e.g.,][]{Muzerolle2000,Demars2025} and in the IR \citep[e.g.,][]{Muzerolle1998c,Natta2004,Salyk2013,Rigliaco2015,Alcala2017}. This UV excess has the capacity to excite and even dissociate some molecules present in the inner disk. Previous observations of solar-mass classical T-Tauri stars (CTTS, $0.2<M_\star<2$) showed that the emission of several molecules is correlated to this accretion luminosity \citep[$\rm H_2O, HCN, C_2H_2,CO_2$;][]{Banzatti2020,Arulanantham2025} and in particular for the CO fundamental band \citep{Herczeg2011,Banzatti2020,Dickson-Vandervelde2025}. Based on the characteristics of the observed CO lines, the CO emission originates from the inner most regions of the disk close to the dust sublimation radius and even partly from the dust free regions of the inner disk where it is truncated by accretion columns. This explains the particularly strong correlation between the accretion luminosity and fundamental CO emission \citep{Najita2003,Bast2011,Banzatti2022}. While the CO-$L_{\rm acc}$ relation has been studied in solar-mass CTTS, such relation has not been studied yet in forming very-low-mass stars (VLMS). However, a different correlation between the stellar luminosity and the $\rm C_2H_2$/$\rm H_2O$ ratio in VLMS \citep{Grant2025} could suggest that the CO-$L_{\rm acc}$ relation behaves differently in VLMS compared to more massive stars.

A dependence between the stellar mass and $ L_{\rm acc}$ (and therefore the stellar accretion rate $\dot{M}_{\rm acc}$) is expected from the magnetospheric accretion process. This correlation has been seen for a large range of stellar masses, despite a notable spread \citep[e.g.,][]{Donehew2011,Manara2015,Rogers2025}. However, the $\dot{M}_{\rm acc}-M$ relation seems to clearly break down at lower masses \citep[$M \lesssim 0.2 M_\odot$][]{Alcala2017,Betti2023,Manara2023,Almendros-Abad2024}, hinting that the accretion process might work differently for these low-mass objects \citep[e.g.,][]{Vorobyov2009,Stamatellos2015,Aoyama2021}. \cite{Aoyama2018} suggested that this change could be due to a change in the accretion shock characteristics for low-mass objects, as the shock is less strong in these objects compared to higher mass objects: when the accretion shock is strong, the emission from the shock is dominated by the preshock while it will be dominated by the postshock otherwise. This change in the shock emission mode may be accompanied by a variation in the UV excess of the accreting star which could impact the molecular emission. Currently, observations show that VLMS might be the key population to understand this break in the $\dot{M}_{\rm acc}-M$ relation, as some of these systems are consistent with the strong shock emission and others with the quiet shock emission \citep{Hashimoto2025}. Characterizing the accretion of VLMS is therefore essential, among others, to then understand if the fundamental CO-$L_{\rm acc}$ relation observed for higher mass stars still holds at lower stellar masses.

Moreover, previous observations have shown that accretion in young stars can be significantly variable \citep[see review by][]{Fischer2023}. This variability can impact the emission of the disk \citep[e.g.,][]{Audard2014,Banzatti2022,Espaillat2023}. Conversely, variable systems constitute unique laboratories to quantify how the emission of the CO fundamental band and of other species like H$_2$O or C$_2$H$_2$ correlates with the accretion luminosity by following over multiple different epochs a single system. JWST observations offer a perfect opportunity to study this variability as it has the required spectral resolution and coverage to detect molecular emission (including CO) and HI emission in the IR simultaneously. The study of emblematic variable systems have already brought first clues on the feedback of accretion onto the disks for Sun-like stars. For DQ Tau, a binary system with modest (factor 4 in $L_{acc}$) but regular bursts, CO lines and, to a lesser extent, hot HCN and H$_2$O, appear to be the most sensitive to the accretion luminosity whereas less excited species remain unaffected \citep{Kospal2025}. In EX Lup \citep{Abraham2009}, the (single) burst was more extreme \citep[factor of more than 50 in $L_{acc}$,][]{Wang2023} and the combination of \textit{Spitzer} and JWST/MIRI observations show a reduction of C$_2$H$_2$, HCN and CO$_2$ and an increase in hot H$_2$O and CO emission during the burst \citep{banzatti_2015,Smith2025}. 

In this paper, we characterize for the first time the variability of accretion of a well-known very-low-mass star, 2MASS J16053215-1933159, which has both HI and CO lines in its spectrum. Our goal is to derive $L_{\rm acc}$ at two different epochs, from new JWST/NIRSpec observations, acquired two years after JWST/MIRI-MRS observations \citep{Tabone2023,Franceschi2024}, and analyze its impact on the emission of the CO fundamental band at each epoch taking advantage of the small ($\sim0.3\mu$m) but critical wavelength overlap between MIRI-MRS and NIRSpec. The present JWST/NIRSpec data also enable to characterize for the first time the full CO fundamental band in a very-low mass star and conduct a deep search for emission from hot organic molecules around $3\mu$m \citep[C$_2$H$_2$, HCN, CH$_4$,][]{Mandell2012}. We start by presenting the observed variability by comparing the new NIRSpec observations with previous MIRI and \textit{Spitzer} observations (Sect. \ref{sec:time_var}), before focusing on characterizing the stellar accretion rate from HI lines at each epoch (Sect. \ref{sec:Macc_Fline}). The evolution of the CO emission is investigated in Sect. \ref{sec:CO_inventory}. We discuss and summarize the global structure of J1605 in Sect.~\ref{sec:disc_struc} and the link between molecular emission and accretion luminosity in Sect.~\ref{sec:disc_CO_acc} before concluding in Sect.~\ref{sec:conclusion}.

\section{Observations}
\label{sec:observations}

\subsection{Our target: J1605}

2MASS J16053215-1933159, called J1605 hereafter, is a very-low mass star known to host a protoplanetary disk. This M4.5 dwarf of mass $M_\star = 0.14 M_\odot$ and luminosity $L_\star = 0.04 L_\odot$ \citep{Luhman2012,Herczeg2014,Carpenter2014} is located in the Upper Scorpius star forming region, at $152.3 \pm 1.27$ pc \citep[EDR3, ][]{Gaia2020} with an age of $2.6\pm1.6$ Myr \citep{Miret-Roig2022}. Its effective temperature is estimated to be around $\sim 3100$ K \citep{Apogee2022,Carpenter2014,Almendros-Abad2024} and its systemic velocity is -3.36 $\rm km.s^{-1}$ \citep{Dahm2012,Apogee2022}. J1605's disk was constrained to be $<0.75M_\oplus$ in dust from the non-detection of the 0.88 mm continuum emission \citep{Barenfeld2016}. Its mid-IR emission is dominated by booming carbon-bearing molecules, with the detection of warm ($\simeq 300-500 K$) $\rm C_2H_2, C_4H_2, C_6H_6, CH_4, HCN$ and $\rm CO_2$ and with little or no water \citep{Pascucci2013,Tabone2023}. 





    





\subsection{Data reduction}
\label{sec:data_red}

\begin{figure*}[t]
        \centering   
        \includegraphics[scale=0.35]{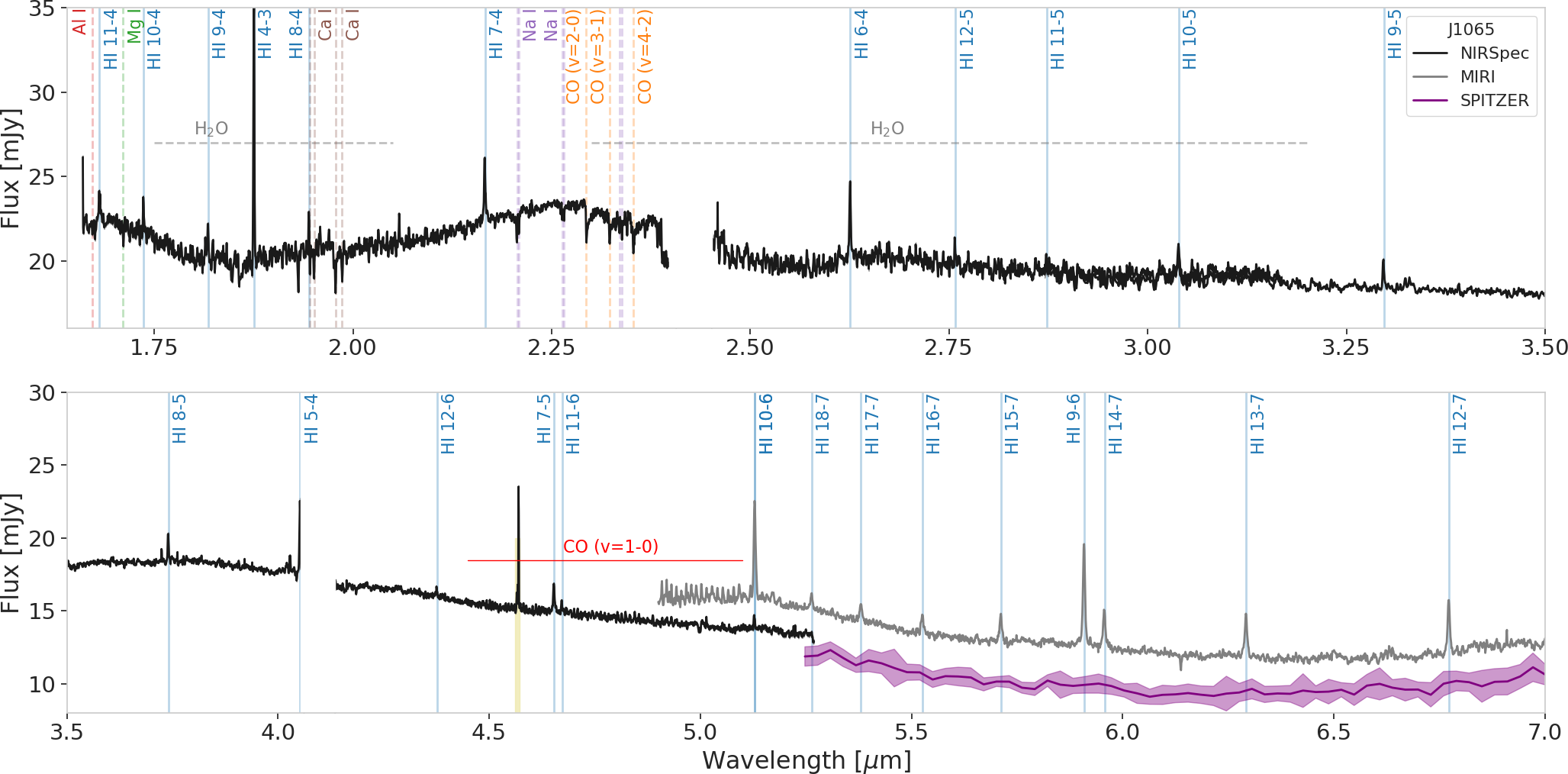}
        \caption{NIRSpec spectrum of J1605 (black solid line) compared to the previous MIRI spectrum (gray solid line) and \textit{Spitzer} spectrum (purple solid line), with their respective error in lighter color. Different HI recombination emission lines are identified with blue vertical solid lines. The CO ($v =1-0$) branches from $\sim 4.4$ to $5.2~\mu m$ are marked in red. Several lines seen in absorption in the stellar photosphere are marked with vertical dashed lines. The yellow shaded area highlights an observational artifact.}
        \label{fig:overview}
\end{figure*}

New JWST/NIRSpec IFU spectroscopy observations \citep{Boker2022,Rigby2023} of J1605 were acquired on the 3rd of August 2024 (program 3962\footnote{\url{https://www.stsci.edu/jwst/phase2-public/3962.pdf}}, PI: B. Tabone). The source was observed at high resolution ($R\sim 2700$)\footnote{For a detailed description of $R(\lambda)$, see \url{https://jwst-docs.stsci.edu/jwst-near-infrared-spectrograph/nirspec-instrumentation/nirspec-dispersers-and-filters}} with the gratings and filters G235H/F170LP and G395H/F290LP ranging in total from $1.66~\mu$m to $5.2 ~\mu$m. While the observations were obtained with a 4-point-dither in case the target showed extended emission, the emission appears, as expected, point-like. The empty sky in the field of view is considered to be sufficient to correct for background contribution. In order to avoid saturation, the \texttt{NRSIRS2RAPID} readout patterns was adopted with 10 groups of integration.

The data was reduced with version 2.0.1 of the pipeline, with the \texttt{jwst$\_$1535.pmap} CRDS map. The MAST default setup was kept except for the \texttt{bkg$\_$subtract} that is skipped as no background data was taken, considering that the empty sky background subtraction was enough when using the \texttt{extract1d} routine of the pipeline. We performed a 3-$\sigma$ clipping at each wavelength. Points that deviated by more than 3-$\sigma$ from the median across the dithers were masked. Afterwards, the dithers were combined into the final spectrum by taking the median of the remaining (non-masked) values at each wavelength. In order to verify whether the observed difference between the MIRI and NIRSpec spectra was an observational artifact or not (see Sect. \ref{sec:time_var}), we extensively investigated the pipeline parameters but found no important data reduction inconsistencies. However, we found that the 1d spectrum extraction in the NIRSpec pipeline is done over a fixed aperture of 0.45", which is not the case in MIRI and could lead to a difference of flux by $<5\%$. Moreover, we found a relation between the \texttt{pixel$\_$replace} parameter and the \texttt{extract1d} routine that still needs to be investigated: when the \texttt{pixel$\_$replace} routine is turned on, the total surface brightness of the extracted spectrum is increased by a few percent while a manual extraction of both spectra does not show such difference. This difference is negligible for our analysis (difference by $<3\%$).



\subsection{Previous MIR observations}


In order to investigate its variability, we gathered previous IR observations of J1605 to compare with the new JWST/NIRSpec spectrum (see Fig. \ref{fig:overview}). \textit{Spitzer} observations were obtained in May 2009 with both the IRS Short-High ($10-19~\mu$m, $R\sim 700$), Short-Low ($5.2-14.5~\mu$m, $R\sim 60-130$) and Long-Low ($14-38~\mu$m, $R\sim 60-130$) modules, as part of the program 50799 (PI: G. Herczeg). A detailed description of the reduction and results can be found in \citealt{Pascucci2013}. J1605 has also been observed by JWST/MIRI with the MRS (Medium Resolution Spectroscopy) on August 2022. This observation was part of the Cycle 1 GTO program 1282 (PI: T. Henning) and made use of the four channel observations available, providing a spectral coverage from 4.9 $\mu$m ($R\sim3500$) to 28.1 $\mu$m ($R\sim1500$). A detailed description of the reduction and results can be found in \cite{Tabone2023}. From both observations, several emission lines from hydrocarbons such as $\rm C_2H_2$ or $\rm C_4H_2$ have been detected. Moreover, several $\rm H_2$ and HI lines have been detected in the MIRI spectrum and were analyzed in \cite{Franceschi2024}.

The WISE telescope surveyed the mid-IR sky in W1 (3.4 $\mu$m), W2 (4.6 $\mu$m), W3 (12 $\mu$m), and W4 (22 $\mu$m) from January-September 2010 in its main mission \citep{wright10} and in W1 and W2 in the extended NEOWISE-R Post-Cryogenic Mission, spanning from September 2013 until July 2024. In each epoch, NEOWISE-R obtained $\sim 13$ individual photometric points across a few days. We obtained single-epoch photometry from the NASA/IPAC Infrared Science Archive (IRSA).  We average the measurements in each epoch to reduce the noise, following procedures developed in \citet{park21} and \citet{contreras23}.  Data points with centroids more than $0.3$ arcsec from the median are excluded.  The W1 and W2 photometry are closely correlated with little color difference between observations, so our analysis focuses on W2.

\section{Results}

\subsection{Time variability from Spitzer, MIRI and NIRSpec spectra}
\label{sec:time_var}

The NIRSpec spectrum is presented in Fig. \ref{fig:overview}. It can be separated in two main parts: the shortest wavelengths ($\lambda < 3~\mu$m) are dominated by the emission from the stellar photosphere (see Appendix \ref{app:stellar_fit}) while longer wavelengths become dominated by the emission from the disk. We identified several atomic hydrogen recombination lines, tracing the accretion of the system, as well as some hints of disk emission from the CO fundamental band which was already detected from the high P-branch lines in MIRI \citep{Tabone2023}. All the other identified features are attributed to absorption from the stellar photosphere.

Notably, while warm $\rm H_2$ lines have been detected in the MIRI spectrum \citep{Tabone2023,Franceschi2024}, the hot $\rm H_2$ rotational and ro-vibrational lines are not detected in the NIRSpec spectral range. The energy of the lines expected between $1~\mu$m and $5~\mu$m corresponds to the highly excited states of the $\rm H_2$ molecule. However, the fit of the lower rotational lines seen in MIRI indicates that the $\rm H_2$ gas is roughly $\rm T=635 \pm 94$ K \citep{Franceschi2024}, which is too cold to excite the hot $\rm H_2$ ro-vibrational lines in NIRSpec. Moreover, we do not detect hydrocarbons as previously detected with MIRI \citep{Tabone2023}. We discuss this non detection in Sect. \ref{sec:disc_hydro}.

\begin{figure}[t]
        \centering   
        \includegraphics[width=8cm]{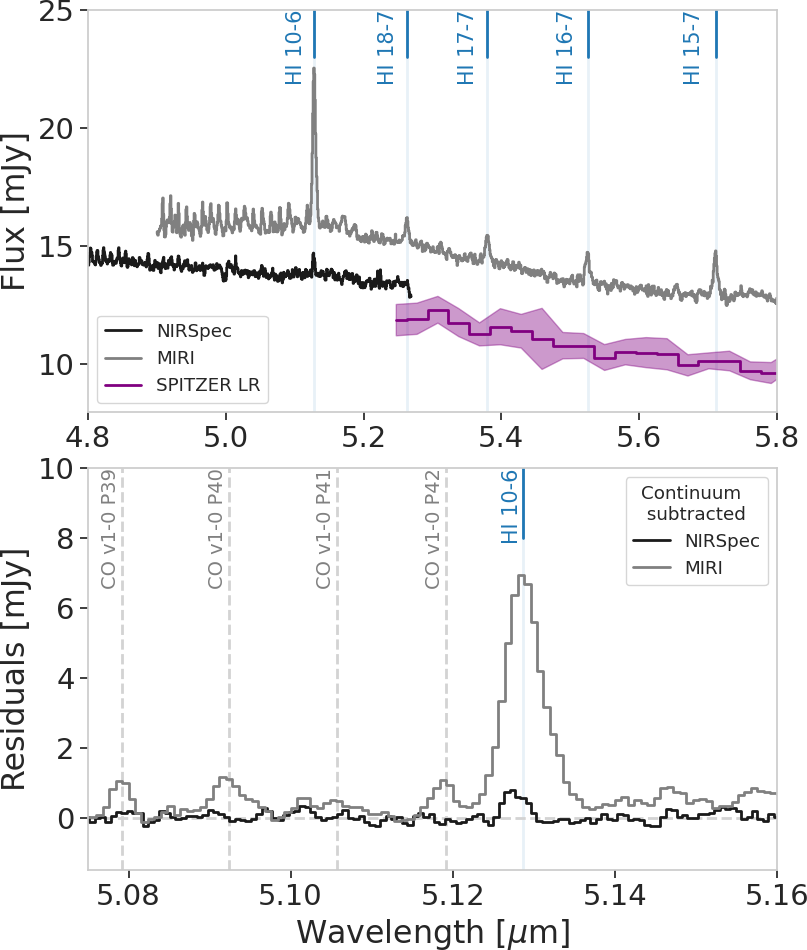}
        \caption{Zoom over the wavelength overlap between Spitzer (2009), MIRI (2022) and NIRSpec (2024) spectra. Between the MIRI and NIRSpec epochs, the continuum flux density got reduced by $\sim 15\%$, the HI 10-6 line is narrower and its flux is reduced by a factor $\sim 13$ (bottom panel) and the fundamental CO emission by a factor $\sim 2$. Here, no correction of the stellar photosphere has been applied.}
        \label{fig:zomm_var}
\end{figure}

NIRSpec and MIRI have a small overlapping wavelength range between $\lambda = 4.9~\mu$m and $\lambda = 5.3~\mu$m, allowing us to probe the evolution of the high P-branch lines of the fundamental CO band and the HI 10-6 transition line. Between the MIRI (2022) and NIRSpec (2024) epochs, the emission of the disk has significantly evolved (see Fig.\ref{fig:zomm_var}): i) the continuum is $\sim15\%$ lower; ii) the HI 10-6 line is roughly two times narrower and $\sim13$ times fainter and iii) the CO emission lines are $\sim3$ times fainter.

\begin{figure}[t]
        \centering   
        \includegraphics[width=8cm]{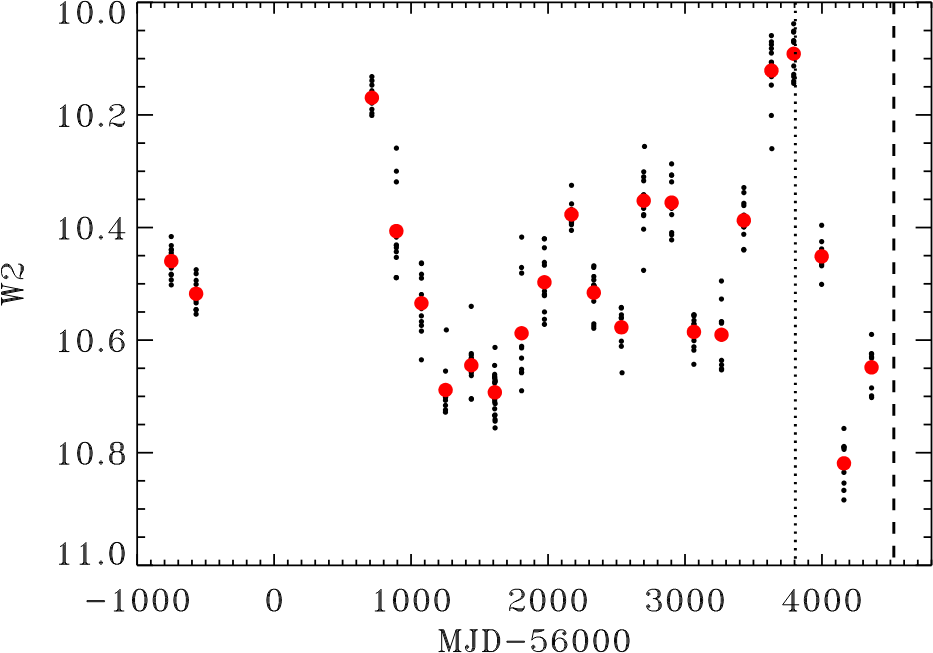}
        \caption{NEOWISE W2 ($\sim4.5~\mu$m) photometry of J1605 as a function of time. The red dots are the averaged values of each measurement. The dotted (resp. dashed) vertical line represents the MIRI (resp. NIRSpec) epoch. The MIRI observation was taken during a peak of luminosity in August 2022 while NIRSpec observation was taken shortly after the end of the monitoring in August 2024.}
        \label{fig:neowise}
\end{figure}

After ruling out the possibility that this change in continuum originates from the data reduction or absolute flux calibration (see Sect. \ref{sec:data_red}), we conclude that these three differences originate from variability in the system. This hypothesis is corroborated by the comparison with the previous \textit{Spitzer} spectrum taken in 2009, where the continuum is significantly lower than both the MIRI and NIRSpec continua, with no clear HI line detection (Fig. \ref{fig:overview} and Fig. \ref{fig:zomm_var}). Variability, in either continuum or line emission, can be generated by different mechanisms (e.g., chromospheric activity, \citealt{Manara2013,Astudillo-Defru2017,Mignon2023}; variable geometry like warps or shadows, \citealt[e.g.,][]{Bouvier2013,Covey2021}; variable extinction along the line of sight \citealt[e.g.,][]{Hillenbrand2013}), however, the flux reduction of the HI line is typical of accretion variability \citep[e.g.,][]{Fischer2023}. 

MIR variability is commonly detected for protoplanetary disks on timescales of days \citep{morales11,leeS24} and years \citep[e.g.][]{rebull2014,park21,2022Zakri}. Fig. \ref{fig:neowise} shows that the NEOWISE W2 brightness of J1605 fluctuates on timescales of years with a standard deviation between epochs of 0.19 mag in W1 and W2, consistent with the 15\% difference in 5$\mu$m continuum emission.  Within each epoch spanning a few days, the average standard deviation is $0.042$ in W1 and $0.046$ in W2.  These short-term changes are dominated by real brightness changes.  This variability is consistent with expectations for typical protoplanetary disk sources.

The W1 and W2 lightcurves do not show any statistically significant period.  A Lomb-Scargle periodogram (\citealt{lomb76,scargle82}, following the implementation by \citealt{contreraspena23}) shows peaks at 530 (strongest power), 152, and 206 days, but all with 10-20\% false alarm probabilities and a Bayesian Inference Criterion of 5, below the level of significance required for a detection.   These values likely overestimate the significance because the data points are not independent.  An analysis of structure functions  (e.g., \citealt{devries05} following methods in \citealt{herczeg23}) for the W1 and W2 lightcurves show a reset timescale of $\sim180$ days.  Although the interpretation of the structure function is somewhat uncertain because the cadence of observations is also $\sim 180$ days, the function establishes that points obtained 1 year or more apart are not correlated.  The timescale for significant mid-IR variability is likely months.

The August 2022 MIRI spectrum was obtained at very close to maximum brightness in the NEOWISE campaign. The August 2024 NIRSpec spectrum was obtained after the final NEOWISE observation of J1605. The previous data points were faint but rising, so the NIRSpec observation is likely close to the median brightness. The difference in continuum emission between these two epochs are large enough to be attributed to the long-term changes rather than the short-term fluctuations.

In order to precisely quantify how the accretion rate evolved and how it impacted the emission of the disk, we start by quantifying the change in stellar accretion rate from the analysis of the HI emission lines (Sect. \ref{sec:Macc_Fline}), before studying the characteristics of the CO gas emission at each epoch (Sect. \ref{sec:CO_inventory}). For each section, we correct the spectra from the contribution of the stellar photosphere, which dominates the emission in particular at the shortest wavelengths of the NIRSpec range. The fit of the stellar photosphere can be found in Appendix \ref{app:stellar_fit}. The correction improves the S/N for the HI line analysis (Sect. \ref{app:HI_lines}) and more details on the impact that it has on the CO analysis is shown in Sect. \ref{sec:stellar_fit}.

\subsection{Variability in accretion rate}
\label{sec:Macc_Fline}

\subsubsection{Inventory and properties of the HI lines}
\label{sec:HI_inventory}

The observations of J1605 show a large variety of HI lines, both with NIRSpec as shown in Fig. \ref{fig:overview} and MIRI-MRS as studied by \cite{Franceschi2024}. For consistency, we re-analyzed the MIRI data and compared our results to \cite{Franceschi2024} in Appendix \ref{app:HI_lines}. In general, we find similar integrated fluxes and report the detection of new faint HI lines.

In Table \ref{tab:HI_lines}, we list all the HI lines identified in both spectra. Each line is fitted with a Gaussian centered on the rest wavelength. To perform the fit, we use the python package \texttt{scipy.optimize.curve$\_$fit}. The fit is applied to the spectrum corrected from the stellar photosphere (see Appendix \ref{app:stellar_fit} and Sect. \ref{sec:stellar_fit}) as it improves the precision of the derived integrated flux by up to a factor 2.

From our fits, we observe that all the NIRSpec lines are blue-shifted by about $-40$  $\rm km.s^{-1}$ on average, much larger than the systemic velocity of the source \citep[$-3.36$  $\rm km.s^{-1}$, ][]{Apogee2022} but smaller than the deconvolved line width (see Appendix \ref{app:HI_lines}). This velocity shift is expected for lines emitted from magnetospheric accretion: depending on the geometry of the system, the lines can present a small redshifted absorption which originates in the capacity of an optically thick disk to shield part of the accretion funnel down to the disk midplane. If this absorption is smaller than or close to the spectral resolution, it will result in an apparent small blueshift of the emission line \citep{Muzerolle2001,Kurosawa2006,Tessore2023}. 

 \begin{figure}[t]
        \centering   
        \includegraphics[width=8cm]{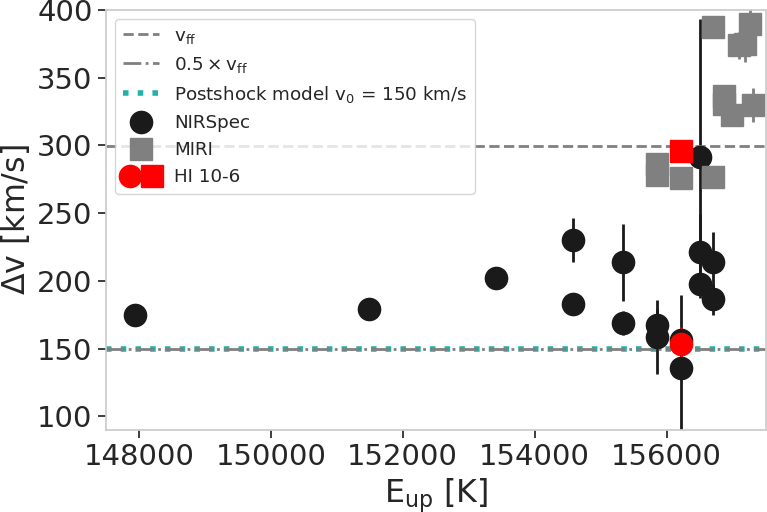}
        \caption{Deconvolved width of the different HI lines as a function of $E_{up}$ from both the NIRSpec (black circles) and MIRI (gray squares) spectra. The HI 10-6 line (shown in red) is observed both in MIRI and NIRSpec and is twice narrower in 2022 than in 2024. The gray dashed line (resp. dotted) shows J1605 free-fall velocity (resp. half of the free-fall velocity). The green dotted line shows the velocity from \cite{Aoyama2018}'s shock model best fit (see Appendix \ref{app:postshock}). }
        \label{fig:HI_width}
\end{figure}

The kinematical broadening of the lines is derived from the fitted line width and is further deconvolved. Following \cite{Marleau2024}, we assume that the point spread function of NIRSpec is a Gaussian, allowing us to derive the deconvolved width $\Delta v_{\rm deconv}^2 = \Delta v^2 - \Delta v_{\rm intr}^2$ where $\Delta v_{\rm intr} = c/R(\lambda)$ is the full-width-half-max width of the instrument based on its resolving power $R(\lambda)$. 
In Fig. \ref{fig:HI_width}, we show the deconvolved width of the lines as function of the energy of the upper level of the transitions. The widths of the lines are independent on the transition and lie around $\Delta v \sim 180$  $\rm km.s^{-1}$, which matches previous observations of VLMS \citep{Mohanty2005}. Moreover, these values are comparable to line widths observed in T-Tauri stars \citep[$\Delta v \sim 200~\rm km.s^{-1}$][]{Muzerolle2001,Wilson2022,Gravity_Wojtczak2023}, which is expected when compared to the free-fall velocity $v_{ff}$ at the surface of a star of radius $R_\star$: 
\begin{equation}
    v_{ff} = \sqrt{\frac{2GM_\star}{R_\star}}
    \label{eq:vff}
\end{equation}
In our case, the free-fall velocity at the surface of J1605 is $v_{ff}~\simeq~300~\rm km.s^{-1}$ (taking $R_\star~=~0.56~R_\odot$ from Appendix \ref{app:stellar_fit} and $M_\star~=~0.14~M_\odot$). For TTS of roughly $R_\star = 3~R_\odot$ and $M_\star = 0.8~M_\odot$, the $v_{ff} \simeq 310 \rm~km.s^{-1}$. Therefore, while TTS are more massive than VLMS, their radius is large enough to make their free-fall velocity comparable to the one of VLMS. Having comparable HI line widths therefore means that the gas accreted by J1605 arrives at the surface of the star with a similar fraction of $v_{ff}$ as TTS.

\subsubsection{Accretion rate from HI lines}
\label{sec:acc_HI_lines}

\begin{table*}[t]            
\centering                          
\caption{Accretion luminosities and rates from different HI lines} 
\begin{tabular}{c c c c c c c}        
\hline             
   Date  & Line & Name & $L_{\rm line}$ & Relation & $L_{\rm acc}$     & $\dot{M}_{\rm acc}$ \\
        &       &      & [$L_{\odot}$] & & [$L_{\odot}$] & [$M_\odot.yr^{-1}$] \\
\hline                 
   2022 & 10-7 & -  & $(1.04\pm0.05)\times10^{-6}$ & Shridharan+2026 & $5.86\times 10^{-2}$ & $(1.18\pm0.00)\times10^{-8}$  \\
    
    2022 & 7-6$^{(*)}$  & -  & $(1.23 \pm 0.14)\times10^{-6}$  & Shridharan+2026 & $5.33\times 10^{-2}$ & $(1.08\pm0.00)\times 10^{-8}$  \\

\hline
    
   2024 & 7-4  & Br$\gamma$ & $(3.60\pm0.22)\times10^{-6}$ & Alcala+2017 & $3.49\times10^{-3}$ & $(7.05\pm0.51)\times10^{-10}$ \\
   2024 & 7-5  & Pf$\beta$  & $(6.99\pm0.55)\times10^{-7}$ & Salyk+2013 & $4.88\times10^{-3}$ & $(9.87\pm0.70)\times10^{-10}$ \\
\hline
  
\end{tabular}

{\raggedright \vspace{3mm} \textbf{Notes}: $^{(*)}$ The HI 7-6 flux is taken from \cite{Franceschi2024} as they properly removed both the $\rm C_2H_2$ and HI 11-8 flux overlapping at this wavelength range}.\\ 

\label{tab:Lacc}
\end{table*}

\begin{figure}[t]
        \centering   
        \includegraphics[width=9cm]{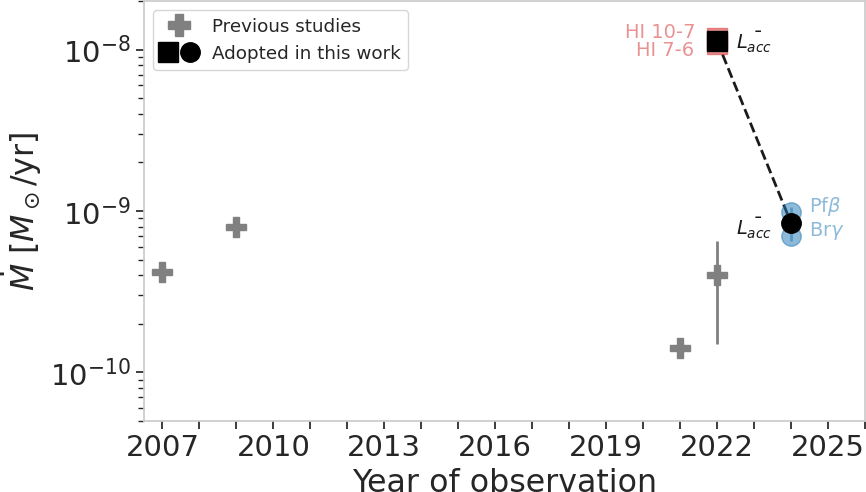}
        \caption{Time evolution of J1605's accretion rate. Previous studies (grey cross symbols) consider different accretion tracers (see Sect. \ref{sec:acc_HI_lines} for more details). The squares (resp. circles) represent the accretion rates derived from this work from the MIRI (resp. NIRSpec) lines. The accretion rate clearly decreased between 2022 and 2024.}
        \label{fig:acc_J1605}
\end{figure}

The accretion rate can be derived observationally from different emission lines and/or the continuum excess generated by the accretion shock at the surface of the object \citep{Gullbring1998,Muzerolle1998c,Herczeg2008,Alcala2017}. Previous observations of J1605 led to the derivation of its accretion rate using either the continuum excess or emission lines: the first accretion rate for our source was derived from observations of the UV accretion continuum excess in 2007 leading to $\dot{M}_{\star,FUV} = 7.9\times 10^{-10}~ M_\odot.yr^{-1}$ \citep{Pascucci2013}; in 2009, the H$\alpha$ line was observed leading to an accretion rate $\dot{M}_{\star,H_\alpha} = 4.16\times 10^{-10}~ M_\odot.yr^{-1}$ \citep{Fang2023}; in 2021, VLT/X-shooter observations allowed for the observation of multiple lines (CaK, H$\delta$, H$\gamma$, H$\beta$, He$\lambda$587nm, H$\alpha$, He$\lambda$667nm, Pa$\gamma$, Pa$\beta$, and Br$\gamma$) leading to an average accretion rate $\dot{M}_{\star, \rm lines } = 1.41\times 10^{-10} ~M_\odot.yr^{-1}$ \citep{Almendros-Abad2024}; and in 2022, the observation of the HI 7-6 line with MIRI lead to the an accretion rate $\dot{M}_{\star,HI_{7-6}} = (4.0\pm 2.5)\times 10^{-10} ~M_\odot.yr^{-1}$ \citep{Franceschi2024}. These values are shown with plus signs on Fig. \ref{fig:acc_J1605}. We discuss the differences in accretion rate originating from different diagnostics in Sect. \ref{sec:disc_HI_acc}.

With NIRSpec and MIRI-MRS, we have access to several IR hydrogen recombination lines. Several studies investigated the link between $L_{\rm line}$ and $L_{\rm acc}$ at these wavelengths \citep[e.g.,][]{Muzerolle1998c,Antoniucci2014,Alcala2017,Rogers2025}. In order to have the broadest view as possible and compare the accretion rate from the NIRSpec epoch and the MIRI epoch, we investigate the $L_{\rm line}-L_{\rm acc}$ relations for all the lines that we have detected. We found relations for 4 transitions relevant for low mass stars: Br$\gamma$ \citep[HI 7-4; $\lambda = 2.17~\mu$m;][]{Alcala2017}, Pf$\beta$ \citep[HI 7-5; $\lambda = 4.65~\mu$m;][]{Salyk2013}, HI 10-7 \citep[$\lambda = 8.76~\mu$m;][]{Shridharan2026} and HI 7-6 \citep[$\lambda = 12.37~\mu$m;][]{Shridharan2026}. Other transitions have been used to derive $L_{\rm line}-L_{\rm acc}$ relations in this wavelength range, such as the Pa$\alpha$ \citep[HI 4-3; $\lambda = 1.88~\mu$m;][]{Rogers2025}, Br$\alpha$ \citep[HI 5-4; $\lambda = 4.05~\mu$m;][]{Testi2025} and Pf$\gamma$ \citep[HI 8-5; $\lambda = 3.74~\mu$m;][]{Testi2025}, however, these studies are based on massive star populations (Herbigs) behaving differently than lower mass stars. In general, we shall stress that the $L_{\rm line}-L_{\rm acc}$ correlations found in the literature are associated with large spreads of one dex and are derived from various stellar population samples. All these relations are discussed in Sect. \ref{sec:disc_HI_acc}. For each $L_{\rm acc}$ derived, we can then estimate the stellar accretion rate following \cite{Gullbring1998}:
\begin{align}
    \dot{M}_{\rm acc} &= \frac{L_{\rm acc}R_\star}{GM_\star} \left( 1 - \frac{R_{\star}}{R_{in}} \right)^{-1} \label{eq:Macc}
\end{align}

\noindent where $R_{in} = 5 R_\star$ is the classical truncation radius when considering magnetospheric accretion \citep[e.g.,][]{Gullbring1998,Herczeg2008,Gravity_Wojtczak2023}. 

In Table \ref{tab:Lacc}, we list all the $L_{\rm acc}$ and resulting $\dot{M}_{\rm acc}$ derived from the extinction corrected spectra (see Appendix \ref{app:stellar_fit}). Similarly to \cite{Almendros-Abad2024}, we use the average of the accretion luminosities to derive the accretion rates at each epoch. By doing so, $\dot{M}_{MIRI} = 1.1\times10^{-8} \rm M_\odot.yr^{-1}$ and $\dot{M}_{NIRSpec} = 8.5\times10^{-10} \rm M_\odot.yr^{-1}$ (resp. black square and dot in Fig. \ref{fig:acc_J1605}), leading to a change of factor $\sim13$ between 2022 (MIRI) and 2024 (NIRSpec). This decrease in the accretion rate is consistent with the observed drop in the HI 10–6 flux; the latter decreases by a factor of $\sim13$, suggesting a comparable reduction of $\dot{M}_{\rm acc}$, given that the $L_{\mathrm{line}}$–$L_{\mathrm{acc}}$ relations are close to linear. We note that using the $L_{\mathrm{HI~7-6}}$–$L_{\mathrm{acc}}$ relation from \cite{Shridharan2026} lead to a value for the accretion luminosity that is coherent with the NIRSpec observation but that is $\sim25$ times more luminous than what was reported in \cite{Franceschi2024}. In their work, they derived $L_{\rm acc}$ with the relation derived by \cite{Rigliaco2015}. This relation was based on \textit{Spitzer} observations which were not able to spectrally resolve the HI 7-6 and the HI 11-8 lines, explaining the difference of factor $\sim25$ observed in our study.

Interestingly, the decrease in accretion rate is associated with a significant reduction in line width. In Fig. \ref{fig:HI_width}, we show that the MIRI-MRS lines are broader than the NIRSpec lines by a factor of about two, with velocities comparable or higher than the free-fall velocity. When we focus on the HI 10-6 line (in red in Fig. \ref{fig:HI_width}), which is seen both by MIRI and NIRSpec, we see that its width is significantly lowered. This suggests that the speed of the shocked accreted material got reduced, either by originating from a different truncated radius in the disk or by a change of accretion mechanism. 


\begin{figure*}[t]
        \centering   
        \includegraphics[width=18cm]{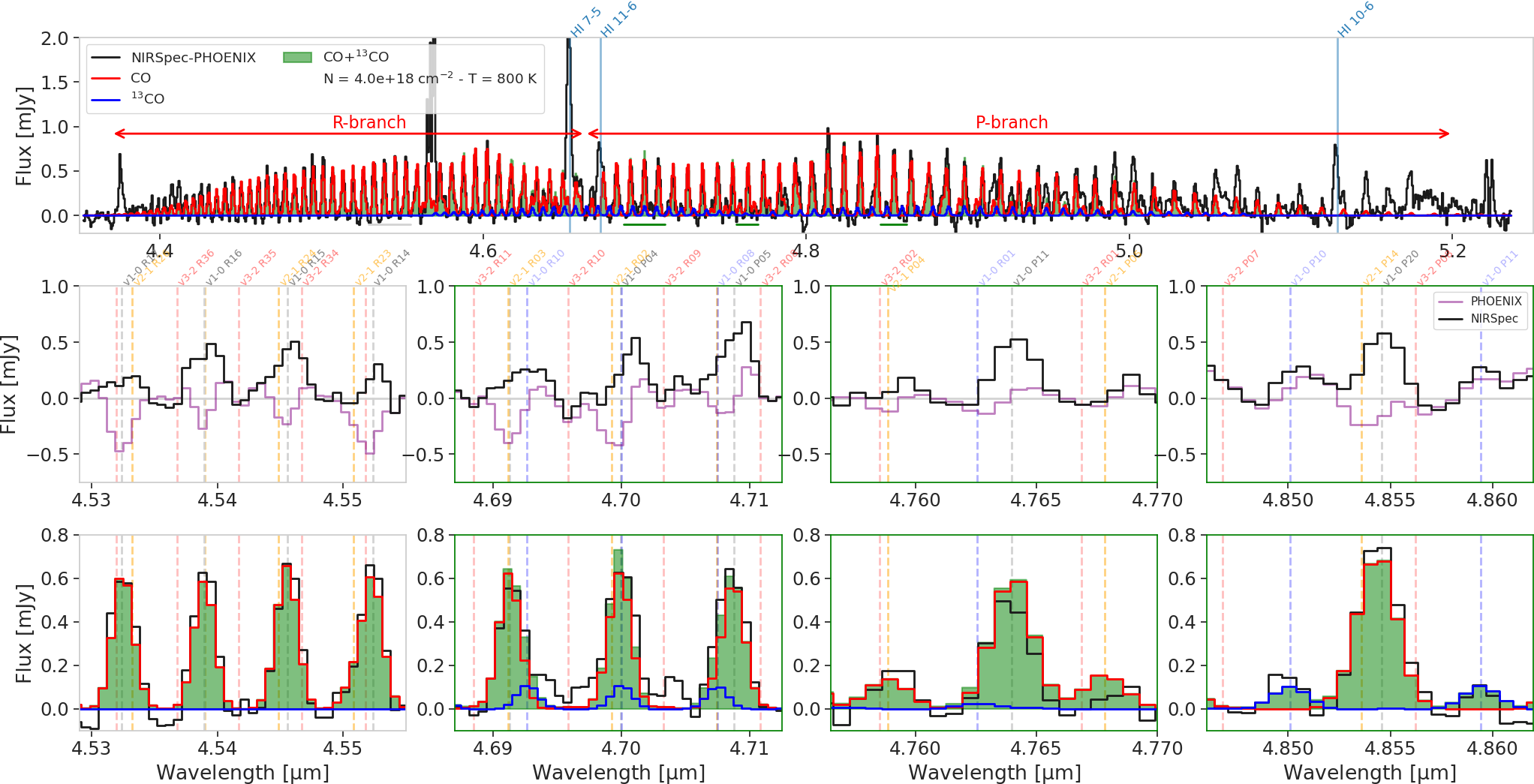}
        \caption{Zoom on CO emission corrected from the photosphere absorption and the continuum. \textit{Top panel:} overview of the corrected CO emission with the best CO slab fit model. The slab model fit was done over three narrow windows marked with the horizontal green lines. \textit{Middle row:} zoom panels of the three narrow windows used for the fit (3 right panels) and of a region of the R-branch (gray horizontal line in top panel) where the CO emission is clearly red-shifted while the stellar photosphere (in purple) is slightly blue-shifted. \textit{Bottom row:} same zoomed in panels as above but with the spectrum corrected of the stellar photosphere, making the CO lines centered. Please note that for readability, we did not label all transitions.}
        \label{fig:zoomCO}
\end{figure*}

\subsection{Variability of CO emission}
\label{sec:CO_diff}
Associated with the drop of the accretion rate by one order of magnitude, the JWST/NIRSpec data also show a reduction of CO emission by about a factor 3. To better characterize CO, an accurate correction for the stellar photosphere is required, in particular for the NIRSpec epoch where the amplitude of CO emission is about that of the photospheric features.

\subsubsection{Impact of the stellar photosphere on the CO emission}
\label{sec:stellar_fit}

The observed CO emission is the result of a combination of the disk emission producing CO lines in emission, and the stellar photosphere producing features in absorption. Therefore, our analysis of the disk emission relies on the robustness of the stellar model which can be well constrained by the short wavelength coverage of the NIRSpec observations (down to $1.66~\mu$m). 
To isolate the emission from the disk at longer wavelengths, we fit our NIRSpec observations with a \texttt{PHOENIX/NewEra} stellar photospheric model\footnote{\url{https://www.fdr.uni-hamburg.de/record/18108}} \citep{Hauschildt2025}. The grid of models is based on the \texttt{PHOENIX/1D} model \citep{Hauschildt1997}, updated with the molecular line data from the \texttt{Exomol} database \citep{Tennyson2016}. The fit of the spectrum includes the contribution from the optically thick inner rim of the disk modelled as a single temperature blackbody. The details of the fit can be found in Appendix \ref{app:stellar_fit}.

The metallicity of the star is a key parameter here as it determines how much CO is present in absorption in the stellar photosphere. With our NIRSpec observations, it is constrained by the absorption of several lines, such as AlI, MgI, CaI, NaI as well as the CO~($v=2-0$), CO ($v=3-1$) and CO~($v=4-2$) overtone lines. All these features, and in particular the CO overtones, are presumably not impacted by the emission from the disk as they trace particularly hot gas ($>10^3-10^4$ K) compared to the expected disk gas temperature ($<10^3$ K, \citealt{Tabone2023}) and they are seen in absorption. Our best-fit model predicts a solar metallicity Fe/H = 0.0, matching the expectation for the majority of the stars in our galaxy \citep{Spina2022,Souto2026}. While the AlI absorption lines seem to be slightly overestimated in our best-fit model, we find that all the other features are correctly reproduced. In particular, the CO overtones ($\simeq 2.3~\mu$m, see Fig. \ref{fig:phoenix_models}) are very well matched by the PHOENIX model, supporting the robustness of our correction for stellar contamination in the CO fundamental band (4.4-5.2 $\mu$m).

\subsubsection{Inventory of CO emission seen by JWST/NIRSpec}
\label{sec:CO_inventory}

\begin{figure*}[t]
        \centering   
        \includegraphics[width=17cm]{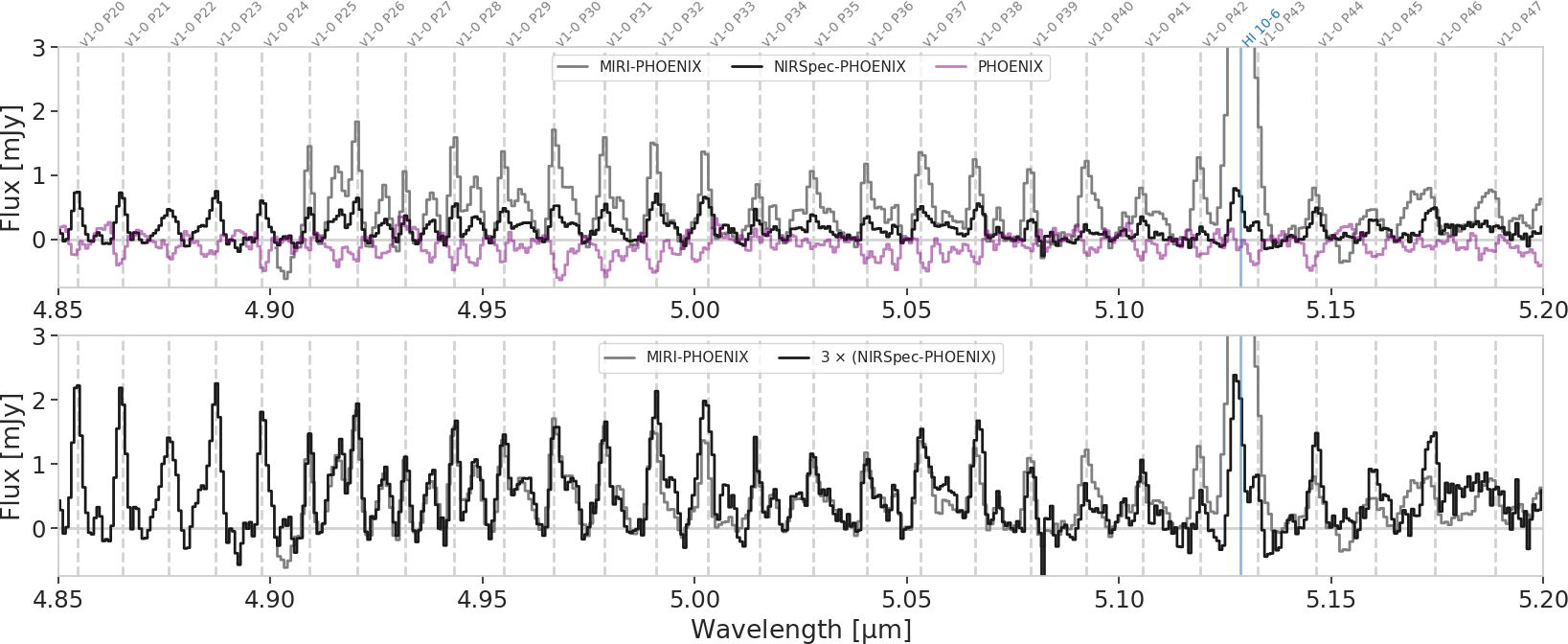}
        \caption{Zoom on the CO P-branch emission from both the MIRI (solid gray line) and corrected NIRSpec (solid black line) spectra. The different CO $(v=1-0)$ are marked with gray vertical dashed lines and the HI 10-6 line is marked with the blue solid line. The remaining lines are CO $(v=2-1)$ lines, which we did not mark for the clarity of the figure. \textit{Top panel:} Direct comparison between the MIRI spectrum, the corrected NIRSpec spectrum and the PHOENIX model (purple solid line). The MIRI spectrum is significantly more luminous than the NIRSpec spectrum. \textit{Bottom panel:} Same as the above panel but the NIRSpec spectrum has been enhanced by a factor 3. The match between the relative P line fluxes shows that the physical characteristics of the CO (column density, temperature) is probably the same at each epochs but emitted over different areas. On the other hand, it is clear that the HI 10-6 line is reduced by a larger factor than the CO emission. }
        \label{fig:overlap}
\end{figure*}

Figure \ref{fig:zoomCO} shows the NIRSpec spectrum subtracted by the best-fit PHOENIX photospheric model. We report for the first time the detection of the full range of CO emission lines belonging to the fundamental band, between 4.4 and 5.2~$\mu$m in a disk around a very-low-mass star. Specifically, the $^{12}$CO($v=1-0$) disk emission lines from $J_{up}=1$ up to $J_{up} =44$ are confidently detected as well as several $^{12}$CO($v=2-1$) and $\rm ^{13}CO$ ($v=1-0$) lines.
The CO disk emission is about the same order of magnitude as the stellar absorption (second row of Fig. \ref{fig:zoomCO}) but thanks to a good fit of the CO overtones around 2.3~$\mu$m, we are confident that the absorption of CO lines is correctly estimated here by our \texttt{PHOENIX} model. Another evidence for the high accuracy of the correction for stellar photosphere can be seen in the position of the prominent CO($v=1-0$). Before correction, there is an apparent shift of the lines (first two panels, middle row, Fig. \ref{fig:zoomCO}). By comparing the CO emission lines to the stellar photospheric model (purple line in the same row), one can note that the photosphere produces absorption not only in the $v=1-0$ transitions but also substantial absorption in the $v=2-1$ and $v=3-2$ transitions. Those excited absorption features distort the cooler lines seen in emission, producing a mismatch in line position. When we correct the NIRSpec spectrum by subtracting our \texttt{PHOENIX} model (bottom row, Fig. \ref{fig:zoomCO}), the CO ($v=1-0$) lines dominating the disk emission become perfectly centered. This confirms the accuracy of the photospheric correction and secures the detection of the aforementioned CO lines. 

We now use the slab model presented in \cite{Tabone2023} and largely used within the MINDS and the JOYS collaborations \citep[e.g.,][]{Grant2023,Perotti2023,vanGelder2024} to derive the properties of the NIRSpec CO emission. The spectroscopic data stem from the HITRAN database \citep{Gordon2022} and include the CO and $^{13}$CO isotopologues. The CO spectrum is further adjusted using a $\chi^2$ method to find the best-fit parameters, namely the column density, effective emitting radius, and temperature. We assume here that $^{13}$CO has the same effective emitting radius and temperature as CO but with a column density $N(^{13}{\rm CO}) = N({\rm CO})/70$, matching the $^{13}$C/$^{12}$C elemental ratio derived from interstellar medium observations \citep{Wilson1999,Milam2005} and found to be consistent with recent protoplanetary disks observations of other species \citep{Salyk2025,Rampinelli2025}. We focus our fit on three specific narrow windows in order to get the best model. The chosen windows are shown with green horizontal lines in the first panel of Fig. \ref{fig:zoomCO}, and panels zooming on each window are shown with green edges just below. They are chosen in regions of the spectrum where we clearly have separated CO ($v=1-0$) and CO ($v=2-1$) emission lines (middle green panel) or isolated $\rm ^{13}CO$ ($v=1-0$) compared to CO lines (right green panel) or a mix of all the three (left green panel) while still scanning a wide range of energy levels (see Table \ref{tab:slab_windows}). From these three windows, the resulting $\chi^2$ map is pretty well constrained (see Fig. \ref{fig:chi2_slab}), leading to a CO emission originating from a rather narrow region of surface $S_{\rm CO} = \pi (0.015 \rm AU)^2$, column density $N({\rm CO}) = 4.0\times 10^{18} ~\rm cm^{-2}$ (and therefore $N(^{13}{\rm CO}) = 5.7\times 10^{16} \rm ~cm^{-2}$) and temperature $T$ = 800~K. 

\subsubsection{Characterization of the variability of CO emission}
\label{sec:CO_evol}

From the corrected spectra obtained by MIRI in 2022 and NIRSpec in 2024, it is now possible to investigate which properties of the CO emitting region varied. Interestingly, our single slab model of the NIRSpec spectrum reproduces well the different rotational transitions of the R-branch ($\Delta J = +1$, and respectively for the P-branch, $\Delta J = -1$) at both short and long wavelengths. It indicates that the bulk CO emission probably originates from a region with a relatively uniform gas temperature of $\sim$ 800~K. The CO emission is also constrained to originate from a compact region almost three times smaller than the $\rm C_2H_2$ emitting region derived from the MIRI-MRS observations of C$_2$H$_2$  ($ S_{\rm C_2H_2} = \pi (0.033 ~\rm AU)^2$, for the optically thick emission at $\sim500$K, \citealt{Tabone2023}). 

While we performed a slab model fit to the MIRI data to compare to the NIRSpec fit, the detection of only a few of the most excited CO lines (from $E_{up,1-0 \rm \,P25} = 4725.0 \rm \, cm^{-1}$ to $E_{up,1-0 \rm \,P47} = 8968.4\rm \, cm^{-1}$) biases the fit towards the highest temperatures as shown in Appendix \ref{app:CO_MIRI} \citep[see also,][]{Grant2024,Francis2024,Kanwar2026}. Even if the best-fit model at MIRI epoch corresponds to hotter slab with lower column density than NIRSpec, the best fit values of $T$ and $N($CO$)$ of NIRSpec lies within the confidence interval. Overall, the fits to the MIRI–MRS and NIRSpec epochs are consistent with a constant temperature and column density, with the flux increase driven solely by an increase in emitting area. This conclusion can be more directly obtained by scaling up the NIRSpec emission by a factor of 3 and compare it to the MIRI spectrum. In Fig. \ref{fig:overlap} (bottom panel), we see that the relative ratio of the different P-branch lines is the same for both epochs, indicating that the CO has a similar column density and temperature in 2022 and 2024 but is likely emitted from a more extended surface at MIRI's epoch (i.e., $\rm S_{CO,MIRI} > S_{CO,NIRSpec}$). 

\section{Discussion}
\label{sec:discussion}

 
\subsection{The dynamical inner disk of J1605}
\label{sec:disc_struc}

\begin{figure}[t]
        \centering   
        \includegraphics[width=9cm]{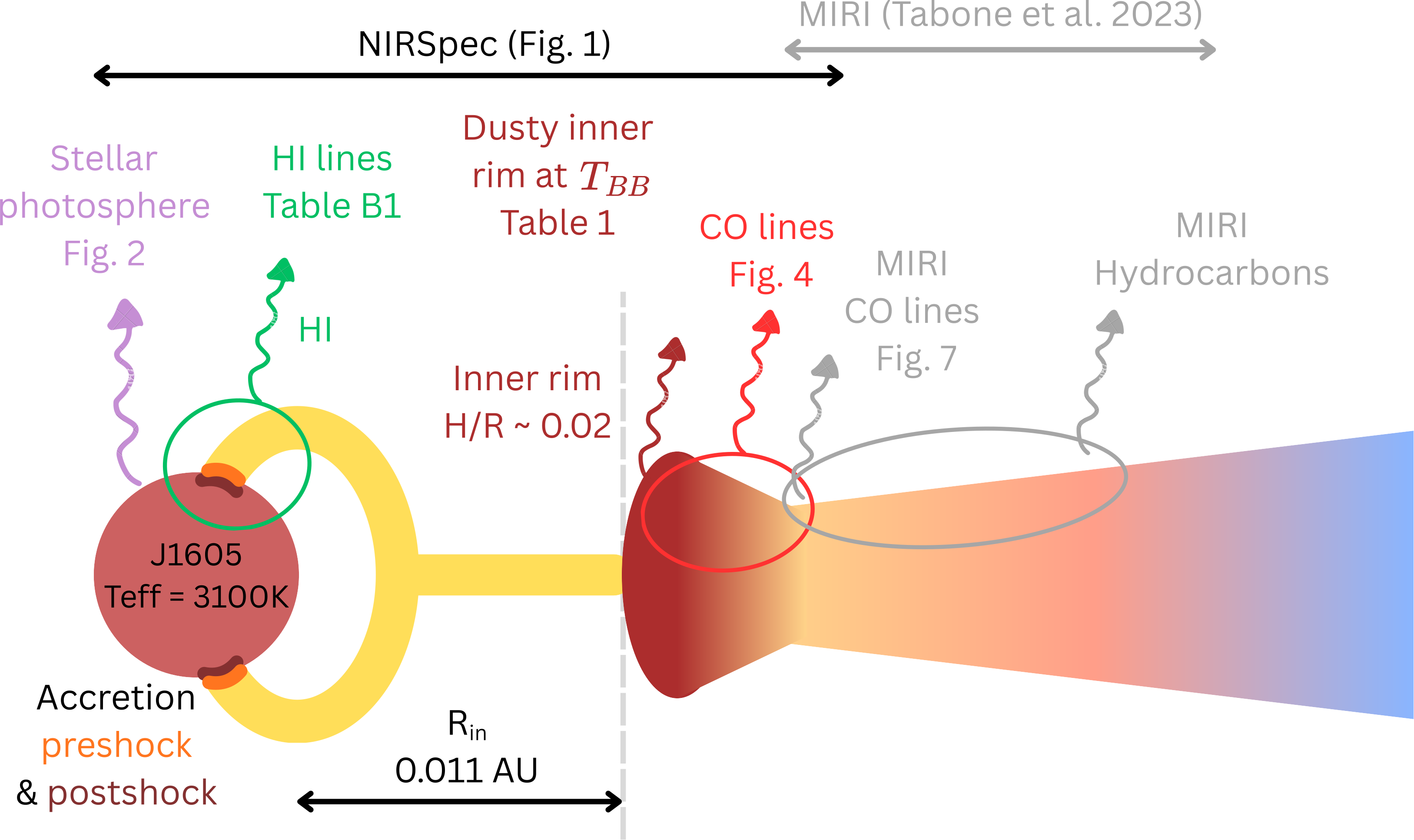}
        \caption{Schematic summary of the global view of J1605. The NIRSpec wavelength range covers the stellar photosphere, emission of the inner rim and the emission from the hottest part of the inner disk, while MIRI's wavelength range allows us to cover the emission from colder parts of the inner disk. The yellow region represent the gas accreted during magnetospheric accretion.}
        \label{fig:schematic}
\end{figure}

The combination of new NIRSpec and archival MIRI-MRS observations allows us to have a unique view of the variable inner disk around J1605. In the previous section, we focused on the variability of the HI and CO emissions. Here, we put together the different components probed across the IR range to build a global view of the inner disk of J1605 summarized in Fig. \ref{fig:schematic}.


\subsubsection{Dust inner rim}

The fit of the near-IR excess reveals the presence of an optically thick dusty inner rim well matched by a single blackbody emission of $T_{in}=1650~\rm K$, close to the sublimation temperature of silicate dust. Therefore, despite the absence of a mid-IR silicate feature, which typically traces warm dust at larger radii, dust is still present at the silicate sublimation front, as in full T Tauri disks. The difference is that, for J1605, this sublimation front is expected to lie much closer to the star, at $R_{in} \simeq~0.01$ AU.

The infered emitting size of the inner rim of $S_{BB} = \pi (0.0067 \, {\rm AU})^2 = 1.41 \times 10^{-4}$ $\rm AU^2$ also provide interesting clues about the structure of the inner rim. To interpret this size, we follow a simple toy model of inner disk emission.
From basic radiative transfer and ignoring the back-heating of the rim, the sublimation radius is expected to be at \citep{Dullemond2001}:
\begin{equation}
    R_{in} = \sqrt{\frac{L_\star}{4\pi T_{in}^4 \sigma_{SB} }},
\end{equation}
where $L_\star$ is the stellar luminosity, $\sigma_{SB}$ the Stefan-Boltzmann constant, and $T_{in} = 1650$ K the temperature derived from our observations. With the luminosity of J1605 ($L_\star = 0.04 L_\odot$) the inner rim is predicted to be located at $R_{in} = 0.011$ AU from the star, as suggested by the chemical modeling of J1605's MIRI spectrum \citep{Kanwar2026}. Regarding the disk scale height at the inner rim, we assume an isothermal disk at $T=1650$ K in hydrostatic equilibrium which gives: 
\begin{equation}
H_{in} = \frac{c_s}{\Omega_K} = \sqrt{\frac{k_BT_{in}R_{in}^3}{\mu m_pGM_\star}}
\end{equation}
where $k_B$ is the Boltzmann constant, $\mu$ the mean molecular weight ($\mu \simeq 2.3$ for a solar composition gas), $m_p$ the proton mass, and $G$ the gravitational constant. From $R_{in}$ and $T_{in}$, $H_{in}$ = 0.00027 AU, leading to an aspect ratio of $H/R = 0.024$. If we assume that the inner rim emits over an annulus of diameter $2 \times R_{in}$ and height $2 \times H_{in}$ we predict an emitting surface of about $S_{in} \simeq 4R_{in}H_{in} \simeq 10^{-5}$ $\rm AU^2$. This surface is a factor 10 smaller than the fitted emitting surface derived from the NIRSpec observations ($S_{BB} = 1.41 \times 10^{-4}$ $\rm AU^2$). The emission of optically thick dust is likely to emit over an altitude that is 2 to 5 times higher than the hydrostatic scale height, bringing our predicted size closer but still below the measured value. Alternatively, a more complex geometry might lead to a radially extended near-IR emission due to e.g. a curved inner rim \citep{Flock2025}. 

Interestingly, we find that the absolute flux at 5$~\mu$m varies only by $15 \%$. Assuming a constant temperature of the inner rim, a variation of the emitting size by the same factor would naturally account for this change in flux. This limited variation in flux suggest that the inner rim of J1605 is relatively stable in stark contrast with the destruction of the inner rim associated with a burst in T Cha \citep{Xie2025} but in line with other variable sources like DQ Tau \citep{Kospal2025} or EX Lup \citep{Abraham2009,banzatti_2015,Wang2023,Smith2025}.

\subsubsection{CO emission as a tracer of the inner disk}

In addition to the narrow spectral region where variability in CO emission is uncovered, our NIRSpec observation also provide an unparalleled characterization of the CO emission covering the full fundamental band. The fit of the CO emission suggests that the emission originates from a relatively compact slab of gas at $T$ = 800~K. This temperature is similar to the CO emission temperature derived in disks around more massive T-Tauri stars \citep[e.g.,][]{Najita2003,Brown2013,Banzatti2017,Banzatti2022}. Thanks to high spectral resolution provided by ground-based spectrometers, the emission in T-Tauri disks is typically separated between a broad component tracing CO inside of or close to the inner rim and a narrow component further out. JWST/NIRSpec cannot spectrally separate the different components but our temperature estimates and the extent of the emission, which is larger than for the dust inner rim, suggest that the bulk CO emission in J1605 comes from further out than the disk sublimation front. In any case, the fact that the derived temperature is close to that infered in T-Tauri disks but the emitting size is much smaller supports the idea that disks around VLMS can be seen as scaled down versions of CTTS.

Interestingly, the detection rate of CO depends on the stellar type and is less frequently detected in disks around VLMS: recent JWST/MIRI observations of disks around VLMS show that only a small fraction of them present CO lines around 5$~\mu$m \citep[e.g.,][]{Arabhavi2025,Morales-Calderon2025}. This difference could simply be due to the combination of a lower accretion luminosity of very-low mass stars, which brings CO line flux close to or below the different instrument detection limits, and their atmosphere producing CO absorption features, hiding the CO emission from the disk as shown in Sect. \ref{sec:CO_inventory} and in \cite{Arabhavi2025}. This last aspect is crucial for future JWST/NIRSpec and ELT/METIS observations of disks around small mass objects (very-low-mass stars, brown dwarfs and even giant planets). These observations will require robust atmosphere fitting in order to properly correct the disk’s emission from the photosphere absorptions.

\subsubsection{Non-detection of near-IR emission of hydrocarbons}
\label{sec:disc_hydro}

\begin{figure*}[t]
        \centering   
        \includegraphics[width=17cm]{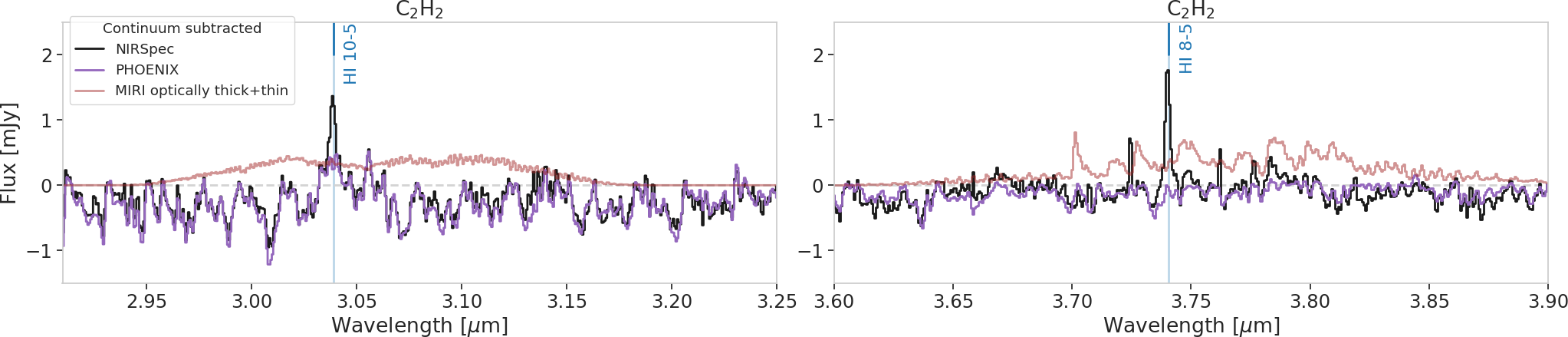}
        \caption{Estimated flux of $\rm C_2H_2$ from its characteristics observed with MIRI assuming LTE emission \citep{Tabone2023}. $\rm C_2H_2$ showed an optically thin emission ($T=400~\rm K$, $N=2.5\times 10^{17}~\rm cm^{-2}$) on top of an optically thick one ($T=525~\rm K$, $N=2.4\times 10^{20}~\rm cm^{-2}$). The emission is expected here to be above the NIRSpec noise level.}
        \label{fig:no_hydro_MIRI}
        \vspace{2mm}
        \includegraphics[width=17cm]{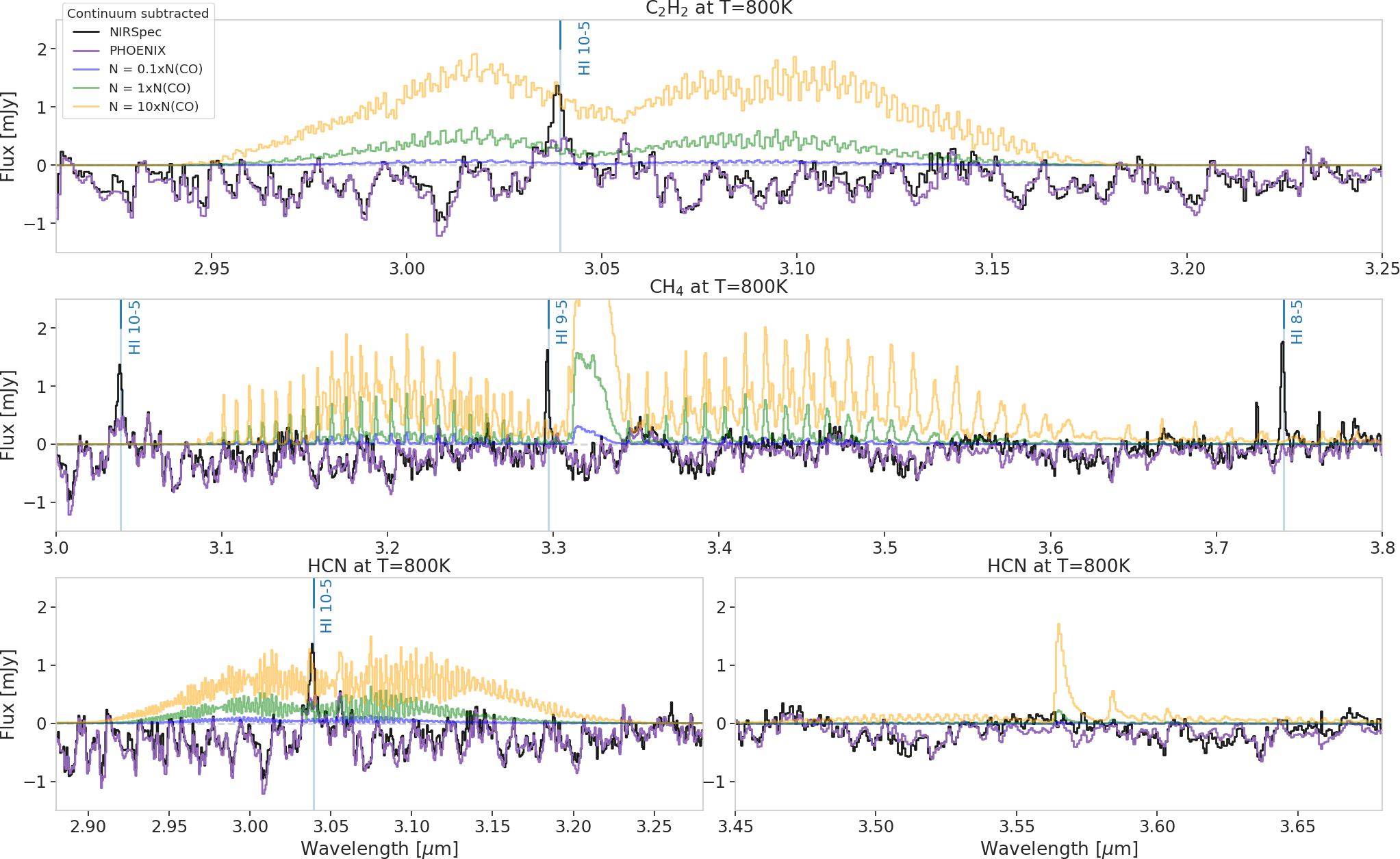}
        \caption{Estimated flux of different hydrocarbons (top: $\rm C_2H_2$, middle: $\rm CH_4$, bottom: HCN) for a hot gas in LTE with the same temperature as observed for CO ($T=$ 800~K) for different column densities based on the CO ($N(\rm CO)=4\times 10^{18}~\rm cm^{-2}$). The non detection of the hot hydrocarbons suggests that the hydrocarbon emission in the warm reservoir is potentially due to non-LTE effects.}
        \label{fig:no_hydro}
\end{figure*}

Previous MIRI-MRS observations of J1605 show that the inner disk is enriched in warm hydrocarbons ($T\sim400-500~$K) with an extremely high column density of observed C$_2$H$_2$ \citep[$N\sim 2 \times 10^{20}~$cm$^{-2}$,][]{Tabone2023}. The emitting region was constrained to be $S_{\rm C_2H_2}=\pi (0.033)$~AU$^2$, which is more extended than both the hot inner rim and the hot CO. The NIRSpec spectral domain covers many features of hydrocarbons emitting from the stretching modes (C$_2$H$_2$, CH$_4$, HCN...). In stark contrast with the MIRI-MRS observation, the NIRSpec spectrum does not show signs of hydrocarbons. Using the same LTE gas slab model that reproduces the prominent MIRI-MRS features of C$_2$H$_2$, signatures of C$_2$H$_2$ should be seen around 3.05 and 3.75~$\mu$m (Fig. \ref{fig:no_hydro_MIRI}). Three reasons can explain this surprising result. First, a dimming of C$_2$H$_2$ emission from MIRI to NIRSpec epoch due to the accretion variability. This would require a reduction of C$_2$H$_2$ near-IR emission by more than a factor of 3, the reduction factor needed to be undetectable by NIRSpec (see Fig. \ref{fig:no_hydro}). This is in tension with the results obtained on outbursting T-Tauri stars \citep{banzatti_2015,Kospal2025} which shows that hot gas traced by CO and excited H$_2$O lines is more sensitive to a change of $L_{\rm acc}$ than warm gas ($\simeq 500~$K). We should expect then to see the hot hydrocarbons gas emission to follow the same evolution as the CO, which is not what we observe. The second possibility is that the reservoir of warm C$_2$H$_2$, that is visible at $> 6~\mu$m, remains hidden by dust opacity around $3~\mu$m. This would require a significant increase in the dust opacity from mid- to near-IR that is difficult to obtain with simple dust models \citep[see e.g., Fig. 3 in][]{woitke_2016}. Moreover, the MIRI's spectrum does not show any evidence for silicate emission meaning that the minimum grain size in J1605 is micronic \citep{Tabone2023,Hyerin2025}. The NIR extinction of the dust is therefore expected to be wavelength independent. The most promising possibility is that large amounts of hydrocarbons are present and emissive in the mid-IR but the density is too low to excite the stretching modes of hydrocarbons. In that case, the stretching modes could be excited by IR pumping enhanced by the near-IR excess \citep{bruderer_2015}. Non-LTE models coupling a realistic disk structure and collisional rate coefficients for the stretching modes are required to fully clarify why the warm reservoir of hydrocarbons is not seen by NIRSpec.

The detection of relatively hot CO testifies the presence of hot gas, which is emissive in the near-IR since the stretching modes of carbon-bearing species like C$_2$H$_2$, CH$_4$, or HCN lie at high energy. Thanks to the characterization of hot CO emission and by assuming LTE emission, the non-detection of these hydrocarbons can be turned into upper limits in their abundance relative to CO. To do so, we compute slab models for C$_2$H$_2$, CH$_4$, and HCN assuming the same emitting area and the same temperature as CO but with column densities ranging from 0.1 up to 10 times that derived for CO. The results shown in Fig. \ref{fig:no_hydro} indicate that, assuming LTE, the abundance of C$_2$H$_2$, CH$_4$ and HCN cannot be larger than about 30\% of CO. While the change of surface emitting area of the CO emission lead to a higher CO flux in MIRI, the column density of hydrocarbons might just be too low in general to be influenced by a similar change of their surface emitting area. In their full thermochemical models of J1605, which match the MIRI-MRS spectrum very well, \citet{Kanwar2026} show that the layer where CO gets abundant has a low abundance of CH$_4$ and even HCN but C$_2$H$_2$ is abundant with a C$_2$H$_2$/CO greater than 1. Provided that non-LTE effects do not impact the near-IR emission of C$_2$H$_2$ in the hot gas, this suggests that the C/O ratio in the gas is not as high as the values suggested by \citet{Kanwar2026} (C/O $\sim~5$ based on the $\rm C_2H_2$ or of $\sim~100$ based on the CO lines), at least in the inner most regions of the disk. Assuming that CO and C$_2$H$_2$ are the main oxygen and carbon carriers, our derived C/O ratio is lower than 1.6. This values is however tentative and detailed thermochemical models building on the pioneering work of \citet{Kanwar2026} are required to obtain robust estimates of the C/O. One of the key outcome of our findings is also that contemporaneous observations with MIRI-MRS and NIRSpec are required to refine the chemical characterization of disks around very-low-mass stars.

\subsection{Robustness of $\dot{M}_{\rm acc}$ derivation at each epoch}
\label{sec:disc_HI_acc}

We find that the accretion rate of J1605 is reduced by a factor $\sim 13$ between 2022 and 2024. However, the accretion rate was derived at each epoch from different HI lines as we relied on two different observations over two different wavelength ranges and no $L_{\rm line}-L_{\rm acc}$ relation was derived for the only HI line seen in both spectra (HI 10-6). We show in Fig \ref{fig:acc_J1605} that the different relations used here are in agreement with each others, giving similar $\dot{M}_{\rm acc}$. These relations are calibrated on stellar populations including low-mass stars: when we consider relations calibrated on populations biased towards massive stars \citep[e.g., mostly including Herbigs,][]{Antoniucci2014,Rogers2025,Testi2025}, we find that the resulting accretion rate is in average 10 times lower than what we derived here. The \cite{Rigliaco2015} relation used in \cite{Franceschi2024} is also calibrated on a too luminous stellar population, leading to the difference in accretion rate that can be seen in Fig. \ref{fig:acc_J1605}. This shows the importance of using $L_{\rm line}-L_{\rm acc}$ relations calibrated on the investigated stellar mass. 

On the other hand, recent studies proposed that the $L_{\rm line}-L_{\rm acc}$ relations derived from  population of T-Tauri stars cannot always be applied to low mass objects: very-low mass stars, brown dwarfs, and planetary-mass companions ($M \lesssim 0.2 M_\odot$). Shock models of \cite{Aoyama2018} suggested that for these low mass objects, the HI emitting region is not the same as for more massive objects. Indeed, they showed that when material is accreted by a low mass object at low speed ($v<200$  $\rm km.s^{-1}$, hydrogen not fully ionized) the emission of the shock is dominated by the emission from the postshock, while the emission of a more violent shock produced by the accretion at high velocity (i.e., stellar accretion with $v>200$  $\rm km.s^{-1}$) is dominated by the emission from the preshock \citep{KF11}. Applying the theoretical $L_{\mathrm{acc}}$–$L_{\mathrm{lines}}$ relations derived for very-low mass objects yields accretion luminosities that are higher by approximately one order of magnitude \citep{Aoyama2021}. VLMS like J1605 have a critical mass where the gas arrives at the stellar surface with velocities $\sim200$  $\rm km.s^{-1}$, meaning that they are sometimes found to be consistent with an accretion shock dominated by the preshock, or by the postshock. 

In particular, \cite{Hashimoto2025} find that J1605's HI emission is consistent with an emission dominated by the postshock, meaning that it would require the use of $L_{\rm line}-L_{\rm acc}$ relations derived for low mass objects rather than the stellar relations mentioned before to determine J1605's accretion rate. Their study was based on X-Shooter observations taken in 2021. One could speculate that the change in HI line luminosity observed from MIRI to NIRSpec epoch is simply a change in HI emitting regime. This would not require a significant change in accretion rate but a small change in the physical properties of the accretion column.

We investigate in Appendix \ref{app:postshock} whether J1605's HI emission is dominated by the postshock or the preshock in 2022 and 2024 by comparing our HI emission to two different shock models (postshock model by \citealt{Aoyama2018}, preshock model by \citealt{KF11}). We find that for both NIRSpec and MIRI epochs, the emission is consistent with a preshock dominated emission and post shock emission models are excluded. Therefore, the change in HI luminosity is unlikely to be due to a change in emission mode and we can apply stellar $L_{\rm line}-L_{\rm acc}$ relations to J1605, at least for the observations taken in 2022 and 2024.

In general, the different $L_{\rm line}-L_{\rm acc}$ relations are roughly linear for both the postshock and preshock relations \citep[e.g.,][]{Alcala2017,Aoyama2021}. Therefore, the relative change of $L_{\rm line}$ compared to $L_{\rm acc}$ is the same no matter the relation used. For J1605, even if no relation has been derived yet for the only transition that is seen by both MIRI and NIRSpec (namely, the HI 10-6), we can expect that a change of total flux of a factor $\sim13$ of this line can be associated with a corresponding change of accretion rate, which is in line with the factor $\sim13$ found from the study of the other HI lines. This is the most robust argument confirming J1605's accretion variability.

\subsection{CO emission as a tracer of $L_{\rm acc}$ variability}
\label{sec:disc_CO_acc}

It has been shown in previous studies that the CO line fluxes scale with the accretion luminosity \citep{Banzatti2017,Dickson-Vandervelde2025}. Our study presents a unique framework to study the correlation between $L_{\rm acc}$ and $L_{\rm CO}$ within the same system: while we see that the stellar accretion rate is reduced by a factor $\sim13$, the CO emission is itself reduced by only a factor of 3. In \cite{Dickson-Vandervelde2025}, they estimate from a population study that the relationship between $L_{\rm CO}$ and $\dot{M}$ can be written as:
\begin{equation}
   \log{(L_{\rm CO})} = (0.54 \pm 0.05) \log{(\dot{M})} -  (0.61\pm0.38) \\
\end{equation}
Therefore, a change of a factor $\sim3$ in CO luminosity should correspond to a change of factor $\sim8$ in accretion rate. This is slightly lower than the change that we observe here, but considering the wide spread in the observed correlation, J1605 still falls in line with their study. In their sample, \cite{Dickson-Vandervelde2025} considered both T-Tauri and Herbig stars. The fact that J1605 matches this $L_{\rm CO}-L_{\rm acc}$ relation is consistent with our previous discussion suggesting that J1605 behaves like a scaled down version of a CTTS.

Other recent studies had the opportunity to follow accretion variability with the CO emission from the disk: in \cite{Kospal2025}, the binarity of the DQ Tau system (2 T-Tauri stars of similar mass $M\sim0.7M_\odot$) creates accretion variability. In this study, the CO flux also does not evolve with a one-on-one correlation with $L_{\rm acc}$ and with a slower power-law than in \cite{Dickson-Vandervelde2025} (change in $L_{\rm CO}$ by a factor $\sim1.5$ for a change of $\dot{M}_{\rm acc}$ by a factor $\sim 1.2$). The hot water component ($\sim 800~\rm K$) seems to be following the same trend as CO (change in $L_{\rm H_2O} $ by a factor $ \sim 1.5$) while the warm water emission ($\sim 650~\rm K$) is almost not impacted by the change of accretion luminosity. These trends show that the emission of CO and hot water is more impacted by the heating by the accretion luminosity than the warm water, hinting that the CO and hot water are located closer to the star and are therefore more sensitive to a change of accretion luminosity. Another recent study by \cite{Almendros-Abad2025} showed that the disk around the planetary-mass object Cha 1107-7626 ($M\sim5-10M_{\rm Jup}$) presents variability consistent with an accretion burst. Even if CO is not detected in this disk, the $\rm H_2O$ emission is completely turned on and off by the accretion variability (reduction of $\rm H_2O$ flux for a change of $\dot{M}$ by a factor $\sim 6-8$). With our study and these previous ones, it is clear that accretion variability is an important aspect to take into account when deriving the composition of inner disks as it can affect the detection and/or total flux of some detected species.

\section{Conclusions}
\label{sec:conclusion}

In this work, we present JWST/NIRSpec observations (August 2024) of a very-low mass star which is compared with previous JWST/MIRI-MRS observations (August 2022). We reveal a dynamical inner disk orbiting J1605 in which a variable accretion induces a change in CO emission. We summarize in Fig. \ref{fig:schematic} all the properties we inferred from the new NIRSpec observations compared to the previous MIRI observations. Our results can be summarized as follows:

\begin{enumerate}    
    \item We report for the first time the detection of the full fundamental R and P branches of the CO from a disk around a very-low-mass star. We detect both the CO $(v=1-0)$, $(v=2-1)$ and $\rm ^{13}CO$ lines. Due to the significant absorption of CO present in the stellar photosphere, we show that it is crucial to correct the spectrum from the central object photosphere. The bulk of CO emission is well reproduced by a single slab model at a temperature of 800~K, typical of the temperature derived in T-Tauri disks. 
    
    \item J1605's accretion rate varied in 2 years between the MIRI (2022) and NIRSpec (2024) observations. The accretion rate was reduced by a factor of $\sim13$. This change of accretion rate, derived from HI line fluxes and in line with the observed reduction of flux by a factor of $\sim13$ of the HI 10-6 line, is supported with a modest reduction of the continuum density flux by 15$\%$ and of the fundamental band of CO flux by a factor $\sim3$. The variation of $L_{\rm CO}$ compared to the variation of $L_{\rm acc}$ is consistent with what is expected for T-Tauri stars, comforting the idea that VLMS can be seen as scaled down versions of CTTS.
    
    \item No hydrocarbon features are detected in the NIRSpec spectrum, despite a satisfactory correction of the photospheric contribution. This contrasts with the booming warm hydrocarbon emission observed with MIRI-MRS. This discrepancy suggests that the hydrocarbon emission in the warm reservoir is muted, potentially due to non-LTE effects. When using CO as a reference and assuming LTE, we find that the C$_2$H$_2$/CO ratio cannot be larger than about 0.3. If C$_2$H$_2$ and CO are the main reservoir of carbon in the CO emitting region, this suggest that the C/O ratio, thought likely elevated, is less extreme than previous thermochemical model found. However, detailed non-LTE disk models are required to bring robust constraints on the C/O ratio in the hot region of J1605's disk. 
    
    \item The HI emission lines are consistent with an accretion emission dominated by the preshock region of the accretion shock at each of the observed epochs, suggesting that J1605 is accreting in a stellar-like accretion regime. This contrasts with previous results for J1605 obtained at a different epoch, which pointed toward a “planetary accretion” regime in which the H I emission is dominated by the post-shock region. The apparent change in accretion rate may be correlated with a reduction in the gas density within the shock region.   
\end{enumerate}

These new NIRSpec observations allowed us to derive new characteristics of the disk structure around a typical very-low-mass star. We show that observations over this spectral range ($1-5~\mu$m) are crucial to precisely characterize the photosphere and be able to extract the disk emission. Future observations (e.g., ELT/METIS) of disks around low mass objects (very-low-mass stars, brown dwarfs, giant planets) will require robust estimates of the central object photosphere if they want to detect the faint CO disk emission hidden in the photospheric absorption. In general, the characterization of disks around low mass objects is crucial to understand the formation of the terrestrial planets and moons observed around such objects but require time-domain observations covering wide range of wavelength.

\begin{acknowledgements}
       This work is based on observations made with the NASA/ESA/CSA James Webb Space Telescope. The data were obtained from the Mikulski Archive for Space Telescopes at the Space Telescope Science Institute, which is operated by the Association of Universities for Research in Astronomy, Inc., under NASA contract NAS 5-03127 for JWST. This research has been funded by the french CNES agency. C. B. C. thanks the following members of the SPACEFORCE group: G.D. Marleau, K. Ward-Duong, J. Bary, C. Rogers, for their useful discussions regarding the accretion processes and their link to HI emission. C.B.C and B.T. thanks also P. Hauschildt for his help with the \texttt{PHOENIX/NewEra} models. A.C.G. acknowledges support from PRIN-MUR 2022 20228JPA3A “The path to star and planet formation in the JWST era (PATH)” funded by NextGeneration EU and by INAF-GoG 2022 “NIR-dark Accretion Outbursts in Massive Young stellar objects (NAOMY)” and Large Gran INAF-2024 “Spectral Key fea-tures of Young stellar objects: Wind-Accretion LinKs Explored in the infraRed (SKYWALKER)”. G.J.H. acknowledges support from National Natural Science Foundation of China general program 12573031 and grant IS23020 from the Beijing Natural Science Foundation.
\end{acknowledgements}

\bibliographystyle{aa}
\bibliography{biblio}

\begin{appendix}

\section{Fitting the stellar atmosphere with \texttt{PHOENIX/NewEra}}
\label{app:stellar_fit}

The spectrum between $1.66~\mu$m $< \lambda < 3~\mu$m is expected to be dominated by the stellar photosphere. In order to isolate the emission from the disk, we fit our NIRSpec observations to a \texttt{PHOENIX/NewEra} stellar photospheric grid of models\footnote{\url{https://www.fdr.uni-hamburg.de/record/18108}} \citep{Hauschildt2025}. This grid of models is based on the \texttt{PHOENIX/1D} model \citep{Hauschildt1997}, updated with the molecular line data from the \texttt{Exomol} database \citep{Tennyson2016}. 

The \texttt{PHOENIX/NewEra} spectra are rebinned using \texttt{SpectRes} \citep{Carnall2017} and convolved to the resolution of both NIRSpec gratings. We use the python function \texttt{scipy.ndimage.gaussian$\_$filter1d} to achieve this goal. However, this function takes as input the array to be filtered and the standard deviation $\sigma$ as a scalar. In our case, $\sigma$ depends on the wavelength as $\sigma = \lambda/(R(\lambda)2\sqrt{2\ln{(2)}})$ with $R(\lambda)$ being the resolving power (which is available on the JWST website\footnote{\url{https://jwst-docs.stsci.edu/jwst-near-infrared-spectrograph/nirspec-instrumentation/nirspec-dispersers-and-filters}} for NIRSpec and which can be found in \cite{Pontoppidan2024} for MIRI). Therefore, instead of having $\sigma(\lambda)$, we look for a wavelength sampling ($\Delta\lambda = \lambda_{i+1}-\lambda_i$) for which $\sigma$ is constant, knowing that $R(\lambda) = \lambda/\Delta\lambda$. 

Assuming that $N$ is the number of spectral elements resolving $\sigma = N/(2\sqrt{2\ln{(2)}})$, then the wavelength sampling follows:

\begin{align}
    \Delta\lambda &= [\lambda_0,\lambda_1,...,\lambda_n] \\ 
    \rm where \; \lambda_{i+1} &= \left(1+\frac{1}{NR(\lambda_i)}\right)\lambda_i
    \label{eq:wave_sampling}
\end{align}
where $\lambda_0$ is the first spectral element of each grating (i.e., $\lambda_0(\rm G235H)$ and $\lambda_0 (\rm G395H)$). $\Delta\lambda$ goes on until $\lambda_{n+1}$ becomes larger than the wavelength range of each grating ($\lambda_{n}(\rm G235H) < \lambda_{max}(\rm G235H)$ and $\lambda_{n}(\rm G395H) < \lambda_{max}(\rm G395H)$)

The \texttt{PHOENIX} model is first resampled to this new wavelength sampling (eq. \ref{eq:wave_sampling}), before being convolved with the gaussian filter with standard deviation $\sigma = N/(2\sqrt{2\ln{(2)}})$. This process is applied to the \texttt{PHOENIX} model for each of our grating (G235H and G395H).

For each model, we consider the emission from the star on top of the emission from the continuum of the inner disk, modeled as a blackbody emission. Based on values found in the literature for J1605, we consider stellar atmospheres of different temperature $T_{eff}$, surface gravity log(g) and metallicity Fe/H. The grid of models explored can be found in Table \ref{tab:phoenix_model}. Moreover, we investigate different disk inner rims characterized with a blackbody temperatures $T_{BB}$ emitting at a radius $R_{disk}$.

\begin{table}[t]            
\centering                          
\caption{\texttt{PHOENIX}, inner rim and extinction model grid used to estimate the contribution from the stellar atmosphere. } 
\begin{tabular}{c c c c}        
\hline             
    Parameter & Range & Step & Best fit J1605 \\
\hline                 
    $T_{eff}$ [K] & 3100 ; 3400 & 100 & 3200  \\
    log(g)        & 2.0 ; 4.0    & 0.5 & 4.0 \\
    Fe/H          & 0.5 ; -1.0 & 0.5 & 0.0 \\
    $k_{scaling}$ & 0.5 ; 1.0 & 0.025  & 0.875 \\
\hline
    $T_{BB}$ [K]  & 1100 ; 1900 & 50  & 1650  \\
    $R_{disk}$ [AU] & 0.004 ; 0.01 & 0.0002 & 0.0067 \\
\hline
    $A_V$         & 0.0 ; 2.0  & 0.1 & 1.0 \\
\hline
\end{tabular}

{\raggedright \vspace{3mm} \textbf{Notes}: $T_{eff}$ corresponds to the star effective temperature, log(g) its surface gravity, Fe/H its metallicity and $k_{scaling}$ a scaling factor of its radius. The inner rim is represented by a blackbody of temperature $T_{BB}$ emitting from an area $S_{BB} = \pi R_{disk}^2$. The extinction from the ISM is represented by $R_V=3.1$ and $A_V$.\\ }

\label{tab:phoenix_model}
\end{table}

Knowing that the radius of an accreting pre-main sequence star differs from the classic Stefan-Boltzmann relation \citep[e.g.,][]{Vorobyov2017}, we consider a scaling factor for the modeled stellar radius such as $R_{\star,model}=k_{scaling}\times R_{\star,SB}$. This scaling factor only impacts the final total flux received.



\begin{figure*}[t]
    \centering   
    \includegraphics[scale=0.35]{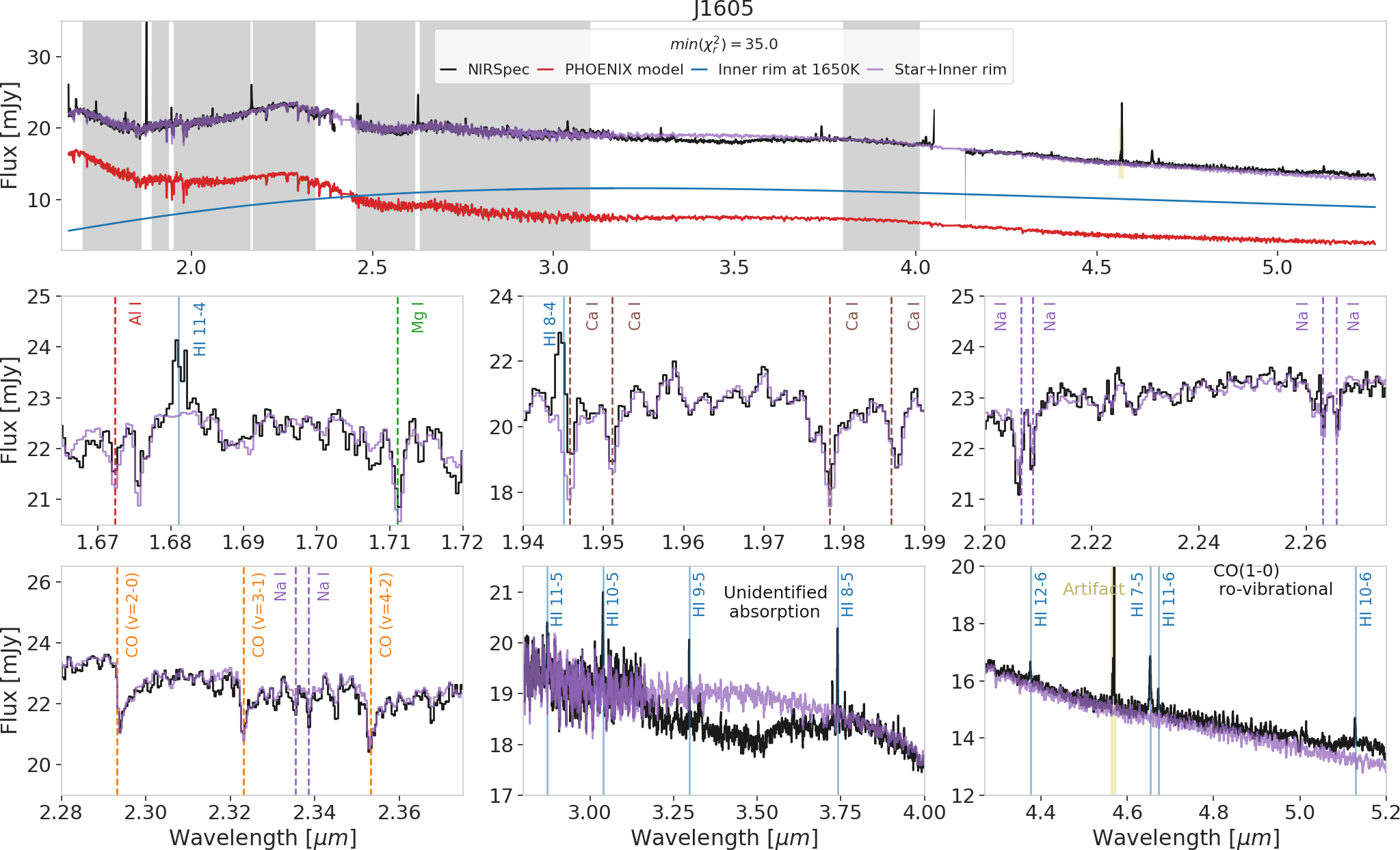}
    \caption{Stellar photosphere best fit model for J1605. The NIRSpec spectrum (black solid lines) is fitted by a stellar atmosphere model (\texttt{PHOENIX}, red solid lines) on top of a continuum blackbody emission originating from the inner rim (blue solid lines). The resulting modeled spectrum (purple solid lines) fits correctly the observations, except for an absorption feature at $3.4 \mu$m. The yellow shaded area highlights an observational artifact, while the gray shaded areas show the regions used for the $\chi^2$ fit. The different panels are zoomed versions of the spectrum over key windows.}
    \label{fig:phoenix_models}
       
\end{figure*}

We calculate the reduced $\chi_r^2$ parameter for the different stellar models and inner rim properties as follows:
\begin{equation}
    \chi^2_r = \frac{1}{N-n} \sum^N_i \frac{(F_{obs,i} - F_{model,i})^2}{\sigma_i^2}
\end{equation}
\noindent where $N$ is the total number of spectral elements and $n$ the degree of freedom (here $n = 7$), and $F_{obs}$ and $F_{model}$ are the observed and modeled flux respectively. Each $\chi_r^2$ based on the simulations grid (Table \ref{tab:phoenix_model}) is calculated considering a dust extinction law modeled by the Python package developed by \cite{Gordon2024} \texttt{dust-extinction}\footnote{\url{https://dust-extinction.readthedocs.io/en/latest/}}. We reproduce the analysis with different extinctions, considering $R(V) = 3.1$ and $A_V$ varying between 0.1 and 2 and we keep the grid giving the lowest $\chi^2_r$.

The fit of the spectrum is conducted on specific spectral windows considered to be negligibly impacted by the emission from the disk (e.g., avoiding bright HI emission lines, see Sect. \ref{sec:HI_inventory}). These regions are marked with the gray shaded areas in Fig. \ref{fig:phoenix_models} where we present the best fit model. The best fit parameters are listed in the last column of Table \ref{tab:phoenix_model}. We find an effective temperature $T_{eff} = 3200 \rm~ K$, which matches the estimates derived by previous studies (3200~K \citealt{Apogee2022},  3100~K \citealt{Carpenter2014} and 3085~K \citealt{Almendros-Abad2024}, 3000~K \citealt{Souto2026}). Moreover, previous observations estimate that J1605 spectral type is between M4.4 \citep{Herczeg2014} and M5-M4.75 \citep{Luhman2012}, corresponding to 2980~K (M5) $<T_{eff}<$ 3160~K (M4) according to \cite{Fang2017}.

Regarding the surface gravity, the value of log(g) = 4.0 is typical of young stellar objects of this mass \citep{Feiden2016,Connelley2026}. The stellar radius is estimated to be $R_\star = 0.875 \times  R_{\star,model} = 0.56 R_\odot$, which is expected for this kind of pre-main sequence star \citep[e.g.,][]{Vorobyov2017} 

The inner disk rim continuum is best represented by a blackbody of $T_{BB}$ = 1650K emitting from a surface area $S_{BB} = \pi (0.0067)^2$ AU. This dusty inner rim is expected to be representative of the region of the disk where the dust sublimates. However, determination of the dust sublimation radius is a very complicated process as it depends on the composition of the grains and on the quantity of gas in the disk \citep[e.g.,][]{Tannirkulam2007,Flock2025}. In general, classical silicate grains are expected to sublimate around $\sim 1500$K \citep[e.g.,][]{Natta2001,Benisty2011}, which is slightly colder than our best fit. 

With the model grid as presented before, we found that we have a better match between the models and the observations when we exclude the region between 3.1 and 3.65 $\mu$m. From our best fit model, a residual absorption feature shows up here (see middle panel of Fig. \ref{fig:phoenix_models}). While previous studies found absorption features in this spectral range (e.g., unidentified aliphatic hydrocarbons, \citealt{Luhman2025}; water ice absorption shifted by disk's geometry, \citealt{Bergner2024}), none of them have either the same broad shape or same center of absorption as our residual. In general, an absorption feature due to the disk itself is unlikely in J1605 as all the other disk features are seen in emission and $A_V$ is small, making it unlikely to have a specific absorption along the line of sight. Such residual could also be due to the combination of stellar spots of different temperatures \citep{PerezPaulino2025} and from a distribution of temperatures from the inner rim. Further investigation is required to explain the origin of this residual feature. 


\section{Inventory of HI lines}
\label{app:HI_lines}

\begin{figure*}[t]
        \centering   
        \includegraphics[width=18cm]{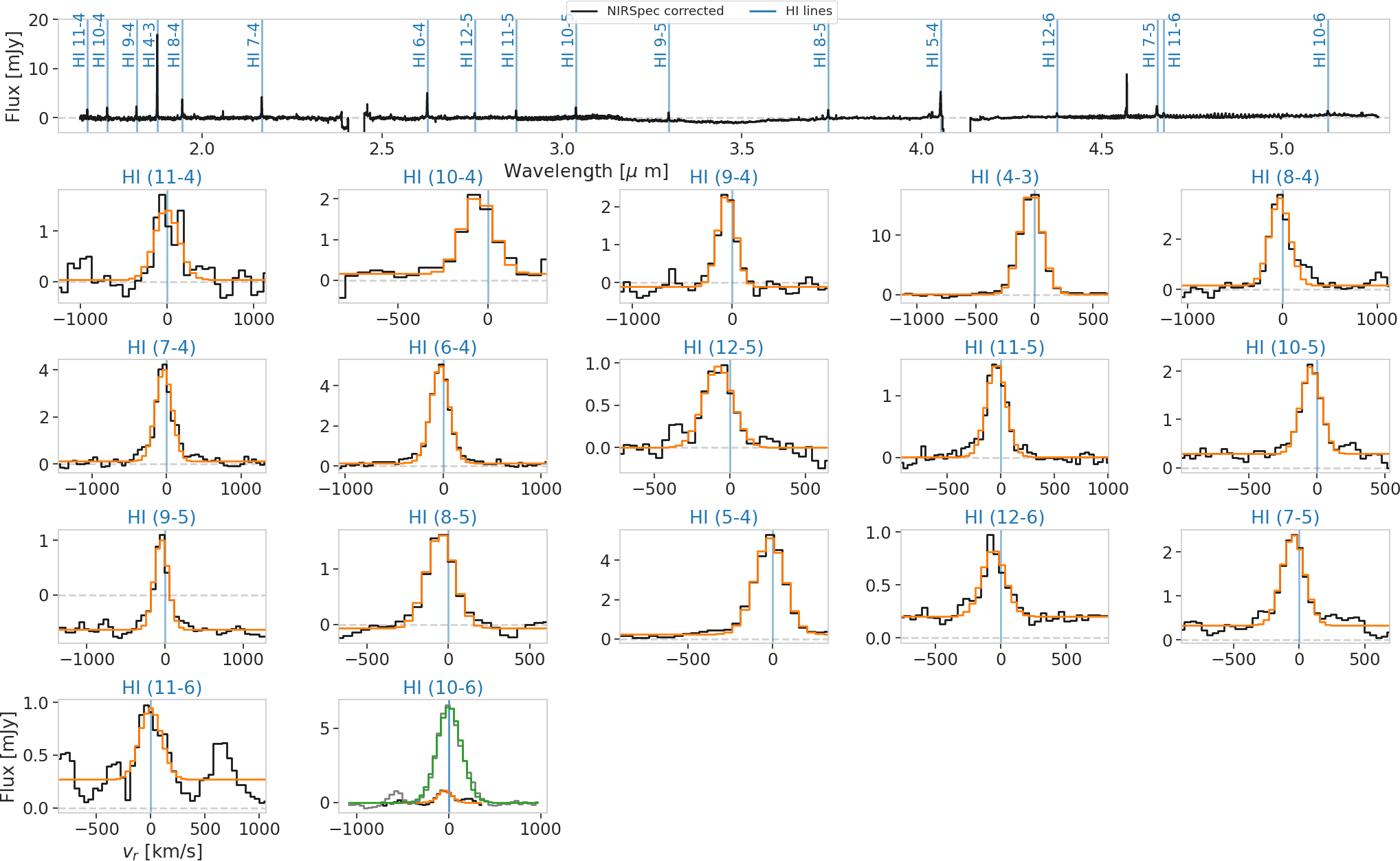}
        \caption{List of all HI identified lines in the NIRSpec spectrum with their gaussian fit in orange. The HI6-10 (Hu $\delta$) line is seen in both NIRSpec (black and orange) and in MIRI (gray and green fit).}
        \label{fig:HI_lines}
                \centering   
        \includegraphics[width=18cm]{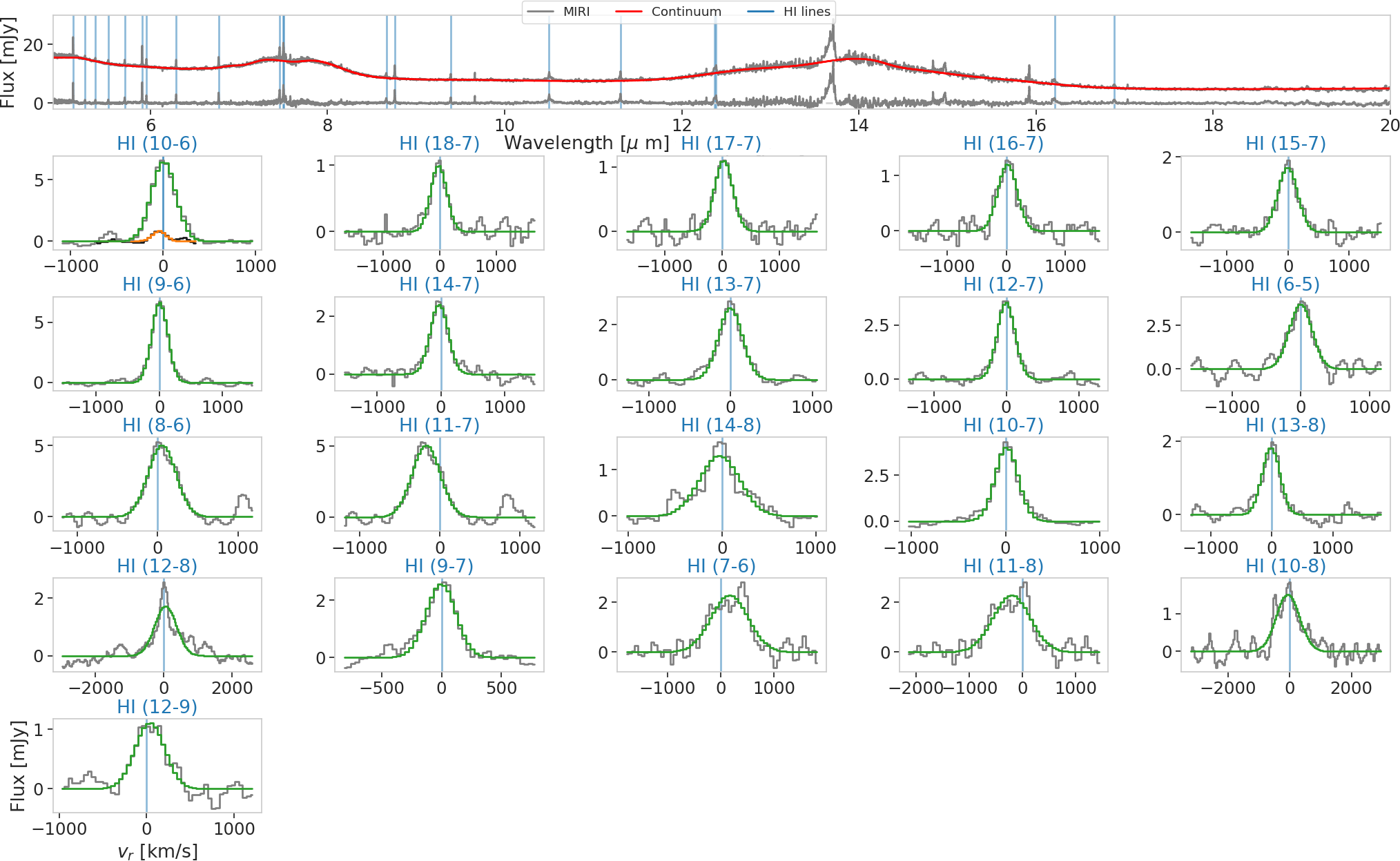}
        \caption{List of all HI identified lines in the MIRI spectrum with their gaussian fit in orange. The HI6-10 (Hu $\delta$) line is seen in both NIRSpec (black and orange) and in MIRI (gray and green fit).}
        \label{fig:HI_lines_MIRI}
\end{figure*}

The fit of the HI and CO lines in the NIRSpec spectrum is done on the corrected spectrum, i.e., after correcting for the ISM extinction and removing the stellar photosphere and inner dust disk black body contributions (see Sect. \ref{sec:stellar_fit}). In Table \ref{tab:HI_lines} we list the differences in integrated fluxes derived from the corrected spectrum (see Fig. \ref{fig:HI_lines}) and compare them to the flux derived from the spectrum only corrected for the extinction. We see that the correction of the stellar photosphere + blackbody contribution gives a better estimate of the line flux thanks to a better estimate of the local continuum corrected from all the small photospheric features.

Similarly to the NIRSpec analysis, the MIRI spectrum is corrected from the contribution of the global continuum to correctly estimate the shape and flux of the HI emission lines. At MIRI wavelengths however, the contributions from the stellar photosphere and inner rim are considered to be negligible for the majority of the spectrum. Based on \cite{Tabone2023}, we interpolate the continuum from different regions of the spectrum considered to be free of important atomic and molecular emission narrow lines (see top panel of Fig. \ref{fig:HI_lines_MIRI}). They showed in their study that $\rm C_2H_2$ is responsible for the two broad continuum emission around $7.5 \mu$m and $14\mu$m. As we are mainly interested in the HI lines of the MIRI spectrum here, we include these two features in the global continuum. However, $\rm C_2H_2$ still produces narrower emission lines, that can overlap with some HI lines, as it is the case for the HI (10-8) transition. 

In Table \ref{tab:HI_lines}, we list all the identified HI lines in our study and compare the integrated flux to \citealt{Franceschi2024}. We report the detection of 5 HI lines that were not identified before: 18-7, 17-7, 16-7, 15-7 and 12-9.  We note that two pairs of transitions are blended with each other and with narrow $\rm C_2H_2$ lines: as we didn't make a detailed study for this case, we took in Appendix \ref{app:postshock} the flux reported in \cite{Franceschi2024} for these transitions. 

\begin{landscape}

\begin{table}[t]            
\centering                          
\caption{Inventory of HI emission lines in both NIRSpec and MIRI} 
\begin{tabular}{c c c c c c c c c c c c}        
\hline             
    Trans. & Name & $\lambda_{0}$ & $A_{i,j}$ & $E_{up}$ & $E_{down}$ & $v_{\rm shift}$  & $\Delta v$ & $\Delta v_{\rm deconv}$  & Integrated flux & Integrated flux & Comment \\
               &      & [$\rm \mu$m] & $\times 10^{4} [\rm s^{-1}]$ & $[\rm cm^{-1}]$ & $[\rm cm^{-1}]$ & [ $\rm km.s^{-1}$] & [ $\rm km.s^{-1}$]    & [ $\rm km.s^{-1}$] & $\times 10^{-16}[\rm erg.s^{-1}.cm^{-2}]$ & $\times 10^{-16}[\rm erg.s^{-1}.cm^{-2}]$ & \\
\hline                 
  NIRSpec & & & & & & & & & & Without correction & \\
\hline
    11-4   &  -              & 1.681 & 2.6  & 906.4  & 6854.8  & $-8.1  \pm 12.9$ & $332.3\pm44$ & $291.9\pm 102$ & $29.40 \pm 6.27$  & $36.61 \pm 8.63$ & - \\
    10-4   &  -              & 1.737 & 4.2  & 1096.8 & 6854.8  & $-55.2 \pm 5.1$ & $204.9\pm18$ & $135.9\pm 54$ & $24.02 \pm 3.43 $ & $37.26 \pm 11.20$ & - \\
    9-4    &  Br$\epsilon$   & 1.818 & 7.5  & 1354.0 & 6854.8  & $-55.8 \pm 2.9$ & $215.7\pm10$ & $158.9\pm 27$ & $31.22 \pm 2.50 $ & $38.55 \pm 8.75$ & -\\
    4-3    &  Pa$\alpha$     & 1.876 & 898.6& 6854.8 & 12186.4 & $-30.22 \pm 0.75$ & $224.9\pm3$ & $175.2\pm 7$ & $218.75 \pm 4.47 $ & $204.41 \pm 6.72$ & - \\
    8-4    &  Br$\delta$     & 1.945 & 14.2 & 1713.7 & 6854.8  & $-29.9 \pm 3.5$ & $253.0\pm12$ & $213.6\pm 28$ & $47.76 \pm 4.17 $ & $27.24 \pm 7.53$ & Im.\\
    7-4    &  Br$\gamma$     & 2.166 & 30.4 & 2238.3 & 6854.8  & $-42.4 \pm 2.1$ & $260.0\pm7$ & $230.3\pm 16$ & $49.67 \pm 2.91 $  & $43.72 \pm 2.77$ & - \\
    6-4    &  Br$\beta$      & 2.626 & 77.1 & 3046.6 & 6854.8  & $-35.6 \pm 0.7$ & $225.0\pm2$ & $202.5\pm 5$  & $44.17 \pm 1.17 $   & $42.10 \pm 2.96$ & - \\
    12-5   &  -              & 2.758 & 1.4  & 761.6  & 4387.1  & $-78.7 \pm 2.9$ & $233.1\pm10$ & $213.6\pm23$ & $8.70 \pm 1.01 $  & $5.93 \pm 2.10$ & - \\
    11-5   &  -              & 2.873 & 2.2  & 906.4  & 4387.1  & $-45.5 \pm 1.4$ & $216.9\pm5$ & $197.7\pm 10$ & $12.11 \pm 0.73 $  & $16.80 \pm 6.33$ & Im.\\
    10-5   &  Pf$\epsilon$   & 3.039 & 3.8  & 1096.8 & 4387.1  & $-44.0 \pm 1.1$ & $177.6\pm4$ & $156.5\pm 8 $ & $11.28 \pm 0.71 $  & $19.36 \pm 2.67$ & - \\
    9-5    &  Pf$\delta$     & 3.297 & 6.9  & 1354.0 & 4387.1  & $-52.2 \pm 1.3$ & $215.2\pm4$ & $167.8\pm 11$ & $11.56 \pm 0.81 $  & $10.25 \pm 1.22$ & - \\
    8-5    &  Pf$\gamma$     & 3.741 & 13.9 & 1713.7 & 4387.1  & $-49.9 \pm 1.0$ & $205.6\pm4$ & $169.1\pm 9$  & $10.01 \pm 0.62 $   & $10.29 \pm 1.22$ & - \\
    5-4    & Br$\alpha$      & 4.052 & 269.9& 4387.1 & 6854.8  & $-17.9 \pm 0.5$ & $208.6\pm2$ & $179.0\pm 4$ & $26.88 \pm 0.88 $   & $24.47 \pm 1.02 $ & Ed. \\
    12-6   &  -              & 4.376 & 1.3  & 761.6  & 3046.6  & $-58.9 \pm 1.5$ & $210.8\pm4$ & $186.4\pm11$ & $3.27 \pm 0.32 $    & $1.71 \pm 0.61$ &- \\
    7-5    &  Pf$\beta$      & 4.654 & 32.5 & 2238.3 & 4387.1  & $-45.7 \pm 1.0$ & $205.1\pm4$ & $183.4\pm 8$ & $9.64 \pm 0.74 $    & $10.22 \pm 1.04$ & - \\
    11-6   &  Hu$\epsilon$   & 4.673 & 2.1  & 906.4  & 3046.6  & $4.7 \pm 3.8$  & $239.7\pm13$ & $221.5\pm 28$ & $3.72 \pm 0.88 $  & - & Blen. CO\\
    10-6   &  Hu$\delta$     & 5.129 & 3.7  & 1096.8 & 3046.6  & $-45.8 \pm 1.6$ & $174.5\pm6$ & $153.6\pm 13$ & $2.99 \pm 0.46 $   & - & Com. \\
\hline
   MIRI & & & & & & & & & & From Franc.2024 &  \\
\hline
    10-6   & Hu$\delta$   & 5.129 & 3.7  & 1096.8 & 3046.6 & $2.6 \pm 0.59$ & $308.5 \pm 2$ & $296.2 \pm 5$  & $41.62 \pm 1.53 $ & $44.6 \pm 2.2$ & Com. \\
    18-7   & -            & 5.234 & 0.1  & 338.5  & 2238.3 & $-32.7 \pm 2.25$ & $345.1 \pm 8$ & $334.1 \pm 16$ & $6.94 \pm 0.77$   & - & \\
    17-7   & -            & 5.380 & 0.2  & 379.5  & 2238.3 & $12.9 \pm 2.27$& $372.5 \pm 8$ & $362.2 \pm 16$ & $8.13 \pm 0.87$   & - & \\
    16-7   & -            & 5.525 & 0.2  & 428.4  & 2238.3 & $10.4 \pm 2.11$& $379.7 \pm 7$ & $369.6 \pm 15$ & $8.79 \pm 0.88$   & - & \\
    15-7   & -            & 5.712 & 0.4  & 487.5  & 2238.3 & $ 11.2 \pm 1.48$& $380.9 \pm 5$ & $370.7 \pm 10$ & $12.26 \pm 0.88$  & - & \\
    9-6    & Hu$\gamma$   & 5.908 & 7.1  & 1354.0 & 3046.6 & $2.7 \pm 0.36$& $299.5 \pm 1$ & $286.3 \pm 3$  & $35.78 \pm 0.83$  & $37.7 \pm 1.3$ & \\
    14-7   & -            & 5.957 & 0.5  & 559.6  & 2238.3 & $-32.2 \pm 0.99$ & $335.6 \pm 3$ & $323.9 \pm 7$  & $14.47 \pm 0.83$  & $17.1 \pm 1.9$ & \\
    13-7   & -            & 6.292 & 0.8  & 649.0  & 2238.3 & $-6.8 \pm 0.71$ & $347.3 \pm 2$ & $335.7 \pm 5$  & $15.36 \pm 0.64$  & $14.9 \pm 1.9$ & \\
    12-7   & -            & 6.772 & 1.2  & 761.6  & 2238.3 & $-5.6 \pm 0.51$ & $291.5 \pm 2$ & $277.3 \pm 4$  & $16.19 \pm 0.62$  & $19.4 \pm 5.3$ & \\
    6-5$^*$& Pf$\alpha$   & 7.460 &102.5 & 3046.6 & 4387.1 & $-10.2 \pm 1.16$ & $388.6 \pm 4$ & $377.6 \pm 8$  & $20.47 \pm 1.49$  & $25.9 \pm 5.8$ & +$\rm C_2H_2$ \\
    8-6$^*$& Hu$\beta$    & 7.503 & 15.6 & 1713.7 & 3046.6 & -               & -             & -              & - & $20.9 \pm 1.0$ & Blen. 11-7 +$\rm C_2H_2$ \\
   11-7$^*$& -            & 7.508 & 2.1  & 906.4  & 2238.3 & -               & -             & -              & - & $32.0 \pm 5.3$ & Blen. 8-6 +$\rm C_2H_2$\\
    14-8   & -            & 8.665 & 0.5  & 559.6  & 1713.7 & $-35.0 \pm 1.79$ & $513.8 \pm 7$ & $505.1 \pm 13$ & $8.25 \pm 0.89$  & $5.3 \pm 1.0$ & WS\\
    10-7   & -            & 8.760 & 3.9  & 1096.8 & 2238.3 & $8.28 \pm 0.42$& $291.3 \pm 1$ & $275.5 \pm 3$  & $14.34 \pm 0.59$ & $15.7 \pm 1.0$ &  \\
    13-8   & -            & 9.392 & 0.8  & 649.0  & 1713.7 & $-16.0 \pm 0.80$ & $348.1 \pm 3$ & $334.5 \pm 5$  & $7.22 \pm 0.49$  & $7.7 \pm 1.0$ &  \\
    12-8   & -            & 10.504& 1.3  & 761.6  & 1713.7 & $2.2 \pm 0.72$ & $732.0 \pm 6$ & $725.2 \pm 12$ & $12.90 \pm 1.02$ & $7.1 \pm 0.9$ & WS \\
    9-7    & -            & 11.309& 8.2  & 1354.0 & 2238.3 & $-6.1 \pm 0.48$ & $296.6 \pm 2$ & $278.6 \pm 4$  & $7.20 \pm 0.44$  & $7.1 \pm 1.0$ & \\
    7-6    & Hu$\alpha$   & 12.372& 45.6 & 2238.3 & 3046.6 & -               & -             & -              & - & - & Blen. 11-8 +$\rm C_2H_2$\\ 
    11-8   & -            & 12.387& 2.3  & 906.4  & 1713.7 & -               & -             & -              & - & - & Blen. 7-6 +$\rm C_2H_2$\\
    14-9   & -            & 12.587& 0.5  & 559.6  & 1354.0 & $29.0 \pm 1.77$& $238.7 \pm 6$ & $213.9 \pm 13$ & $3.38 \pm 1.01$  & - & \\
    13-9   & -            & 14.183& 0.8  & 649.0  & 1354.0 & $-191.9 \pm 2.27$& $396.4 \pm 9$ & $380.5 \pm 19$ & $5.94 \pm 1.88$  & - & \\
   10-8$^*$& -            & 16.209& 4.7  & 1096.8 & 1713.7 & $-104.6 \pm 2.03$& $853.9 \pm 7$ & $845.6 \pm 14$ & $8.50 \pm 1.08$  & $4.5 \pm 1.3$ & +$\rm C_2H_2$ \\
    12-9   & -            & 16.881& 1.4  & 761.6  & 1354.0 & $31.8 \pm 0.60$& $406.8 \pm 3$ & $388.1 \pm 6$  & $2.86 \pm 0.32$  & - &  \\
\hline

\end{tabular}

{\raggedright \vspace{3mm} \textbf{Notes}: Transitions marked with $^*$ have been found to be blended with $\rm C_2H_2$ emission in \cite{Franceschi2024}. Comments: Im. = Improved: the flux has been significantly improved when we correct from the stellar photosphere thanks to a better estimate of the local continuum}; Ed. = Edge: the transition is located close to the NIRSpec wavelength cut-off; Blen. X = Blended with X: the line is blended with X; Com. = Common: the line is seen in both MIRI and NIRSpec; WS = Weird Shape.\\ 

\label{tab:HI_lines}
\end{table}

\end{landscape}

\clearpage

\section{CO analysis of MIRI spectrum}
\label{app:CO_MIRI}

\begin{figure*}[t]
        \centering   
        \includegraphics[width=18cm]{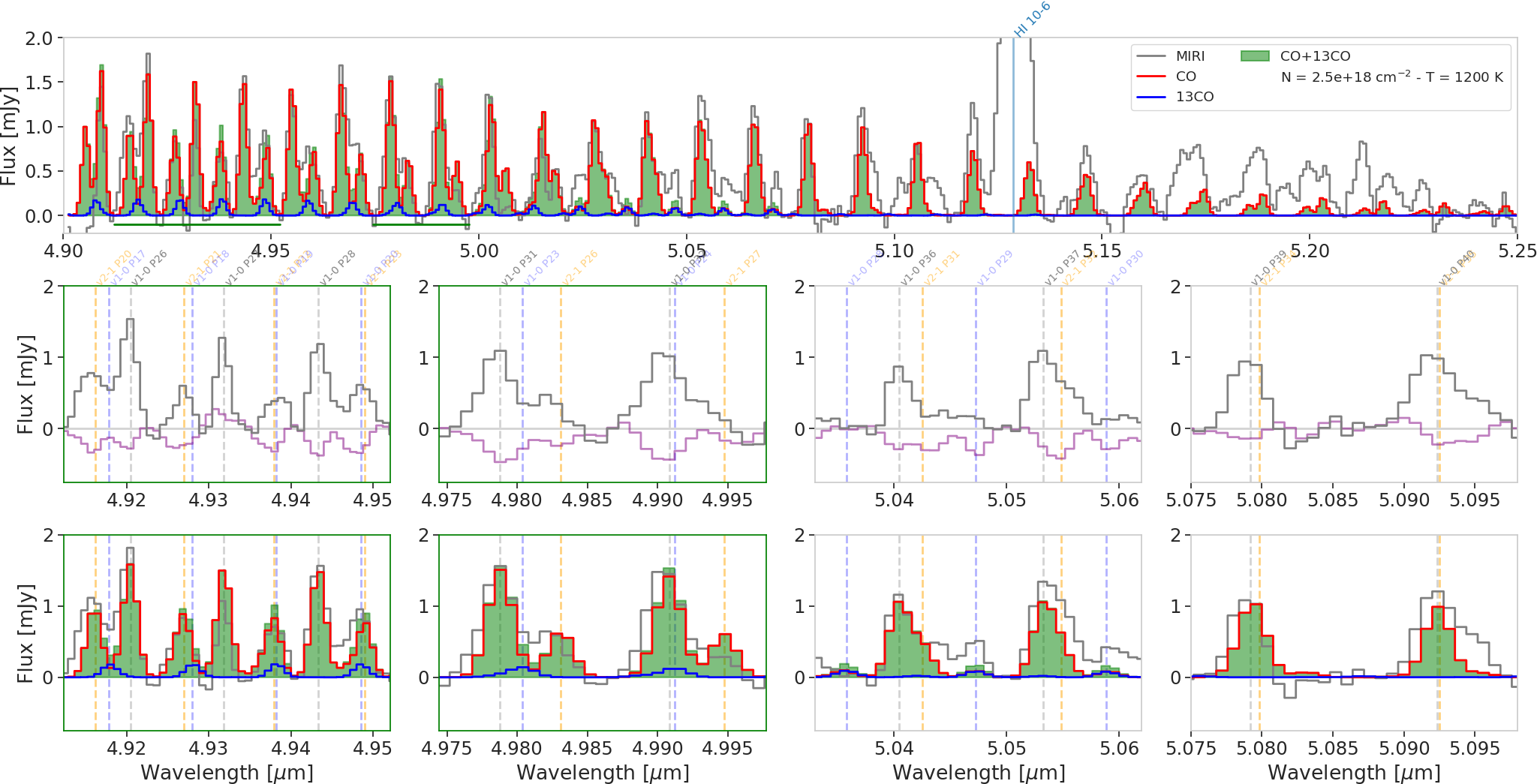}
        \caption{Zoom on CO emission corrected from the photosphere absorption. \textit{Top panel:} overview of the corrected CO emission with the best CO slab fit model. The slab model fit was done over two narrow windows marked with the horizontal green lines. \textit{Middle row:} zoom panels of the two narrow windows used for the fit (2 left panels) and of two more excited regions of the P-branch. \textit{Bottom row:} same zoomed in panels as above but with the spectrum corrected of the stellar photosphere. Please note that for readability, we did not label all transitions.}
        \label{fig:CO_fit_MIRI}
\end{figure*}

\begin{figure*}[t]
    \centering
    \begin{minipage}{.45\textwidth}
        \centering
        \includegraphics[width=6cm]{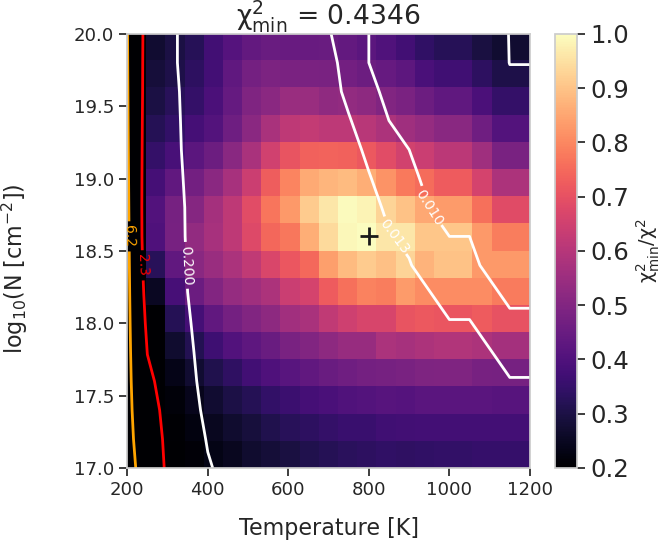}
         \label{fig:chi2_slab}
    \end{minipage}%
    \begin{minipage}{0.45\textwidth}
        \centering
        \includegraphics[width=6cm]{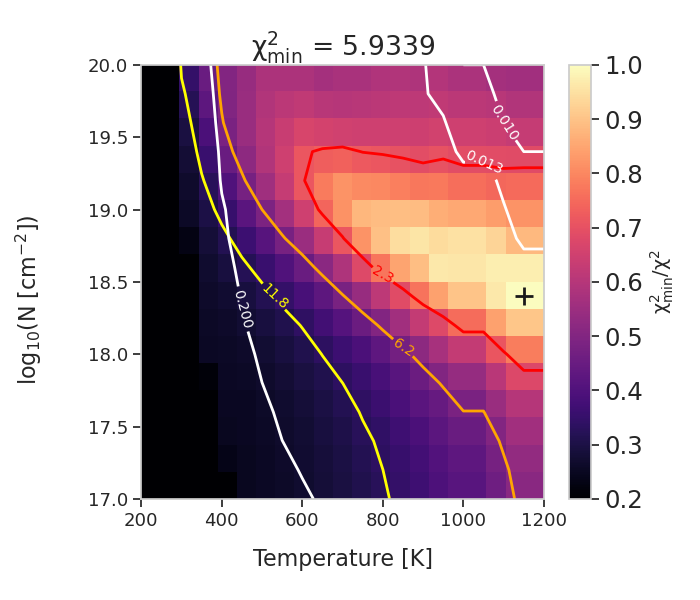}
    \end{minipage}
    \caption{$\chi^2$ maps of CO slab model for NIRSpec (left) and MIRI (right). The different white lines show different emitting radius ($\rm S_{CO} = \pi R_{emit}^2$)}
\end{figure*}

In Sect. \ref{sec:CO_inventory}, we fit the CO emission with a slab model at NIRSpec's epoch as its wavelength range covers the full extent of the CO fundamental band. The specific windows used for this fit are listed in Table \ref{tab:slab_windows} and the $\chi^2$ map of the fits is shown in the left panel of Fig. \ref{fig:chi2_slab}. However, with MIRI's wavelength range, only the high P-branch lines are detected, which biases the fit towards the highest temperature. In Fig. \ref{fig:CO_fit_MIRI}, we show our fit of the CO emission from the MIRI spectrum corrected of the stellar photosphere, as in Sect. \ref{sec:CO_inventory}. As expected, we find that the best fit slab model provides a high temperature of 1200~K for a column density $N({\rm CO})=2.5\times 10^{18}~\rm cm ^{-2})$ emitted over a surface $S_{\rm CO,MIRI}=\pi (0.015~\rm AU)^2$. As this fit is biased, we consider that comparing the relative intensity of the different MIRI P-branch lines to the NIRSpec ones is a more reliable approach to determine the evolution of the CO properties (Sect. \ref{sec:CO_evol}).

\begin{table}[t]            
\centering                          
\caption{Properties of the lines used for the slab fit model} 
\begin{tabular}{c c c}        
\hline             
    Transition & Wavelength & $E_{up}/k_B$ \\
               & [$\mu$m]  & [K] \\
\hline                 
    & Window 1 & \\
\hline
     CO$(v=1-0)$ P05 & 4.70877 & 3138.5 \\
     CO$(v=1-0)$ P04 & 4.69995 & 3116.6 \\
     CO$(v=1-0)$ P03 & 4.69124 & 3100.1 \\

     CO$(v=2-1)$ R01 & 4.70742 & 6145.5 \\
     CO$(v=2-1)$ R02 & 4.69922 & 6161.4 \\
     CO$(v=2-1)$ R03 & 4.69114 & 6183.6 \\

     $\rm ^{13}CO$ $(v=1-0)$ R08 & 4.70746 & 3204.4 \\
     $\rm ^{13}CO$ $(v=1-0)$ R09 & 4.69999 & 3251.6 \\
     $\rm ^{13}CO$ $(v=1-0)$ R10 & 4.69262 & 3304.0 \\
\hline                 
    & Window 2 & \\
\hline
     CO$(v=1-0)$ P11 & 4.76399 & 3385.1 \\

     CO$(v=2-1)$ P05 & 4.76782 & 6183.6 \\
     CO$(v=2-1)$ P04 & 4.75886 & 6161.8 \\

     $\rm ^{13}CO$ $(v=1-0)$ R01 & 4.762562 & 3021.0 \\
\hline
    & Window 3 & \\
\hline
     CO$(v=1-0)$ P20 & 4.85457 & 4123.9 \\

     CO$(v=2-1)$ P14 & 4.85359 & 6623.2 \\

     $\rm ^{13}CO$ $(v=1-0)$ P11 & 4.85942 & 3251.6 \\
     $\rm ^{13}CO$ $(v=1-0)$ P10 & 4.85008 & 3204.4 \\
\hline

\end{tabular}
\label{tab:slab_windows}
\end{table}

\section{Discriminating the postshock and preshock models}
\label{app:postshock}

As mentioned in Sect. \ref{sec:disc_HI_acc}, J1605 have been shown to have an HI emission in 2021 consistent with a postshock dominated emission, unlike more massive stars \citep{Hashimoto2025}. They find that J1605's accretion rate is $1.26 \times 10^{-10}~M_\odot.\rm yr^{-1}$, which is higher than what would be expected if the HI emission was dominated by the preshock ($\dot{M}_{\rm acc} = 6.31\times 10^{-11}M_\odot.\rm yr^{-1}$ when derived from \cite{Alcala2017} $L_{\rm line}-L_{\rm acc}$ relations). Due to J1605's variability, we investigate here the origin of the HI emission lines at each epoch in order to determine which $L_{\rm line}-L_{\rm acc}$ relation is relevant.

Our HI emission lines are compared to the two following models: the preshock model, from \cite{KF11}, matches the HI line ratio observed in T Tauri and estimates the hydrogen line ratios for shocks at different temperatures and densities; the postshock model is from \cite{Aoyama2018} and estimates the line ratios for shocks at different speeds and densities. Both models include the prediction of the hydrogen line emission for the transitions from n = 2 to n = 6. Some previous studies used specific line ratios to disentangle which model fit better the observations \citep{Betti2022,Aoyama2024}. However, these studies do not take into account all the ratios self consistently. Therefore we start by considering all the lines simultaneously before comparing the result to the approach considering specific line ratios.

\subsection{Fit of all the listed HI lines}

\begin{figure*}[t]
        \centering   
        \includegraphics[width=14.5cm]{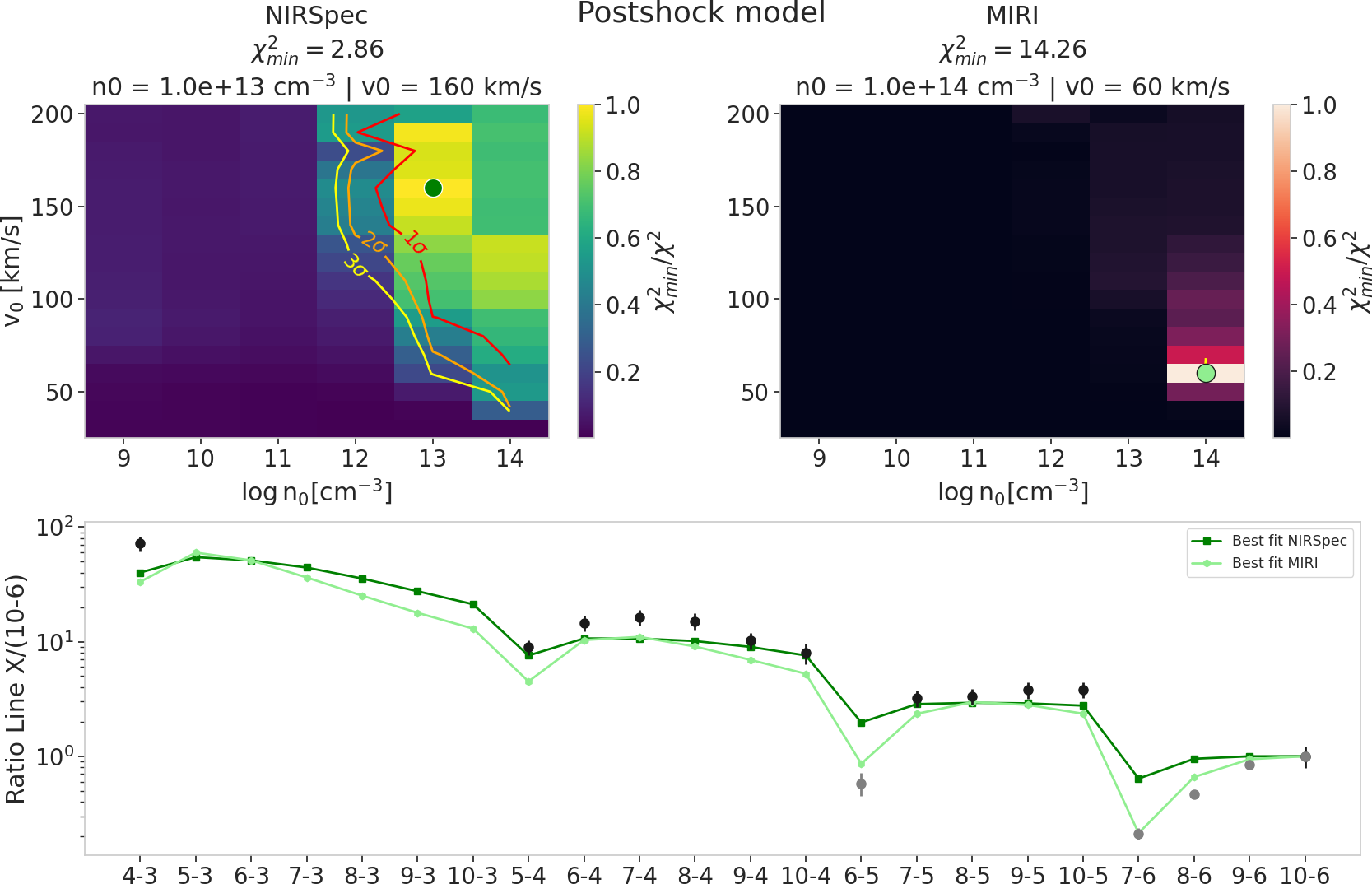}
        \caption{ \textit{Top row:} $\chi^2$ maps for both the NIRSpec (left) and MIRI (right) lines fit with the \cite{Aoyama2018} model. The best fit model is shown with a green dot and the different lines show the estimate of the error of the fit with $\chi^2-\chi^2_{min} = [2.3,6.2,11.8] = [1,2,3]\sigma$. \textit{Bottom panel:} Line ratios for the different lines in both the NIRSpec (black dots) and MIRI (gray dots) spectra, with their corresponding best fit model (dark green line for NIRSpec and lime line for MIRI).}
        \label{fig:chi2_shock_models_post}
\end{figure*}

\begin{figure*}[t]
        \centering   
        \includegraphics[width=14.5cm]{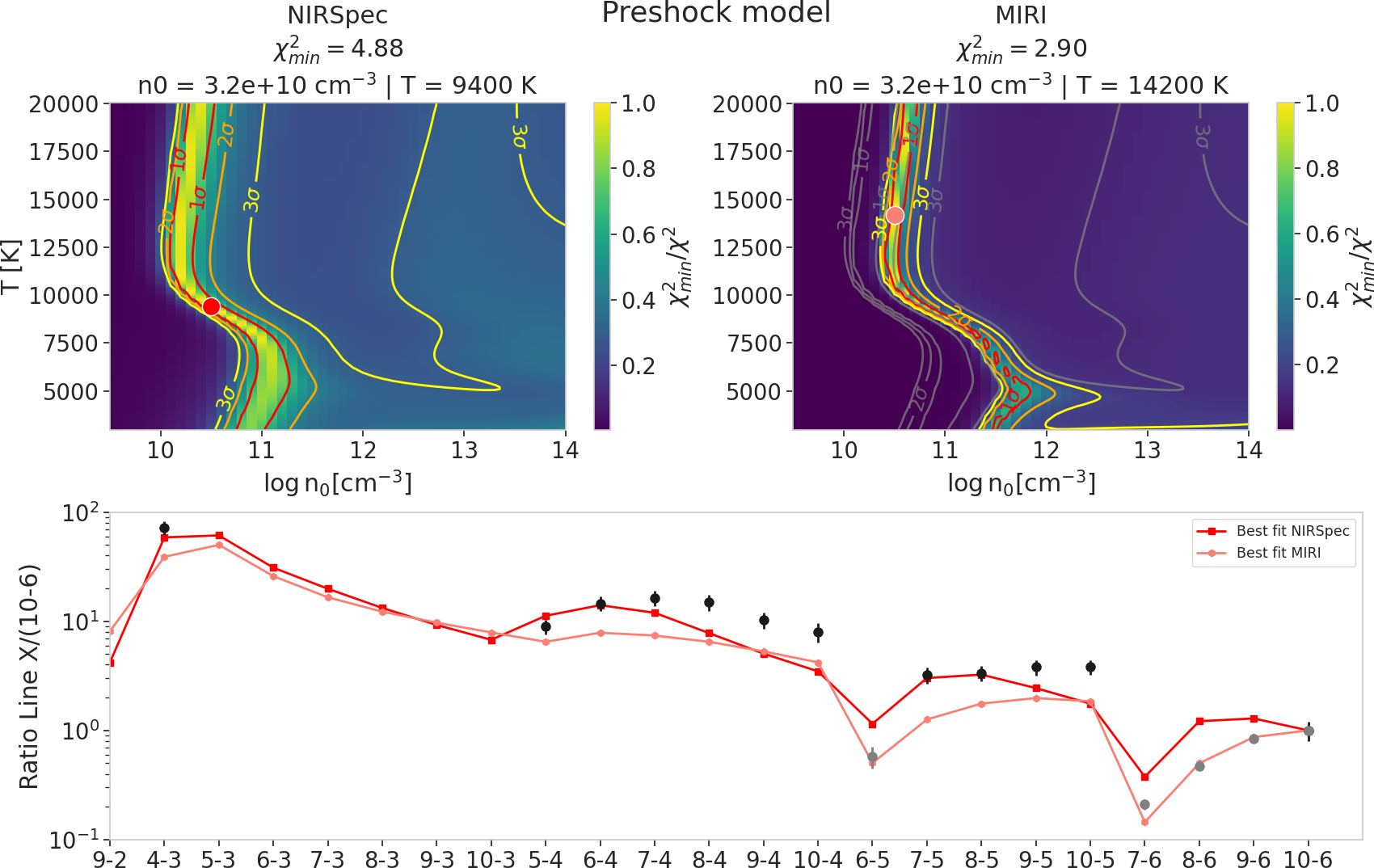}
        \caption{Same as Fig. \ref{fig:chi2_shock_models_pre} but for the fit to the \cite{KF11} model. The gray lines from the right panel show the different $\chi^2$ contours from the NIRSpec epoch for comparison. The density is systematically lower at NIRSpec's epoch.}
        \label{fig:chi2_shock_models_pre}
\end{figure*}

We perform a $\chi^2$ analysis to fit the different line ratios for each epoch: for this, we consider the 12 (resp. 5) detected lines in NIRSpec (resp. MIRI) included in the models. We show our best fits for the postshock model in Fig. \ref{fig:chi2_shock_models_post} and for the preshock model in Fig. \ref{fig:chi2_shock_models_pre}. The confidence intervals are determined from $\Delta \chi^2 = \chi^2 - \chi_{min}^2$ following \cite{Avni1976} and \cite{Press1992} (see Table p.815 for a 2 parameter fit): the $1\sigma$ interval corresponds to $\Delta \chi^2 = 2.3$, $2 \sigma$ to $\Delta \chi^2 = 6.2$ and $3 \sigma$ to $\Delta \chi^2 = 11.8$.

For MIRI's epoch, it is pretty clear that the emission of the HI lines is dominated by the preshock: the "best" fit from the postshock model has a high $\chi^2\sim14$ and it predicts a very low speed $v_0 = 60 {\rm~km.s^{-1}} \simeq0.2v_{ff}$, which is too low for magnetospheric accretion. The preshock model on the other hand, gives a good fit to the observations: $\chi^2 \sim 2.9$ for a shock with density $n=3.2\times10^{10} \rm ~ cm^{-3}$ of temperature $T = 14 200~\rm K$. While we see from the $\chi^2$ map that the temperature is poorly constrained in this model, we find similar results as in \cite{Franceschi2024}. The authors used only one ratio (HI (9-7)/HI (7-6)) to compare to the \cite{KF11} model, and they constrain the density to be between $7\times10^9 ~\rm cm^{-3}$ and $4 \times 10^{10} ~\rm cm^{-3}$ but could not constrain the temperature. Previous studies of T-Tauri stars show that these values for the density and temperature of shock are typical of T-Tauri low accretors \citep{Antoniucci2017,Campbell2023}. Interestingly, it has been shown also that these preshock parameters are not significantly influenced by the stellar type \citep{Saad2024}.

For the NIRSpec's epoch, the results are less striking: both models give reasonable fits, with maybe a slightly better match for the postshock model. Both models are consistent with realistic properties with $v_0 = 160~\rm km.s^{-1}$ and $n_0 = 10^{13}\rm~cm^{-3}$ for the postshock model and $T=9400~\rm K$ and $n_0 = 3.2\times10^{10}\rm~cm^{-3}$ for the preshock model. Again, the temperature is poorly constrained here. We can see that both best-fits give systematically different densities, with values up to $10^3$ times lower in the preshock case than in the postshock one. This difference might originate from the fact that both models trace different parts of the accretion flow which have different properties. 

In order to determine if the emission could really originate from the postshock during the NIRSpec epoch, we can derive the resulting accretion rate from the best-fit parameters ($n_0$ and $v_0$) and compare it to the accretion rate determined from the single line emission (see Sect. \ref{sec:Macc_Fline}). Based on \cite{Aoyama2018} and assuming that the extinction is negligible, the accretion rate can be written as follows:
\begin{equation}
    \dot{M} = 4 \pi \mu v_0 n_0 d^2\frac{F_{obs}}{F_{mod}}
\end{equation}
where ($v_0$, $n_0$) are the best fit shock velocity and density, $d$ the distance of the system, $F_{obs}$ the observed integrated flux from a given HI line and $F_{mod}$ the integrated flux of this line as predicted by the postshock model for our best-fit parameters. When applying this to the measured flux listed in Table \ref{tab:HI_lines} and considering $R_\star = 0.56 R_\odot$ and d = 152 pc, we find an accretion rate $\dot{M} \sim 10^{-8} M_\odot.yr^{-1}$ for the NIRSpec epoch, which is comparable to the accretion rate derived for the MIRI epoch, despite the notable decrease of the HI10-6 line (see Sect.\ref{sec:Macc_Fline}). We therefore conclude that the emission from the accretion shock in J1605 for at least these two epochs is dominated by the preshock, excluding the possibility to be dominated by the postshock.

Even if the shock is probably dominated by the preshock at both epochs, we still find an evolution of the shock parameters. In Fig. \ref{fig:chi2_shock_models_post}, we over-plotted the different confidence intervals from the NIRSpec $\chi^2$ map onto the MIRI $\chi^2$ map (left panel onto right panel) to compare how the fitted parameters evolved between each epoch. Even if the temperature is poorly constrained in both cases, the NIRSpec shock is systematically at a lower density than the MIRI shock (by a factor $\sim 10$). This is consistent with the decrease of accretion luminosity derived from the individual emission lines (Fig. \ref{fig:acc_J1605}), confirming that the emission from the shock is dominated by the preshock. 


\begin{figure}[t]
        \centering   
        \includegraphics[width=9cm]{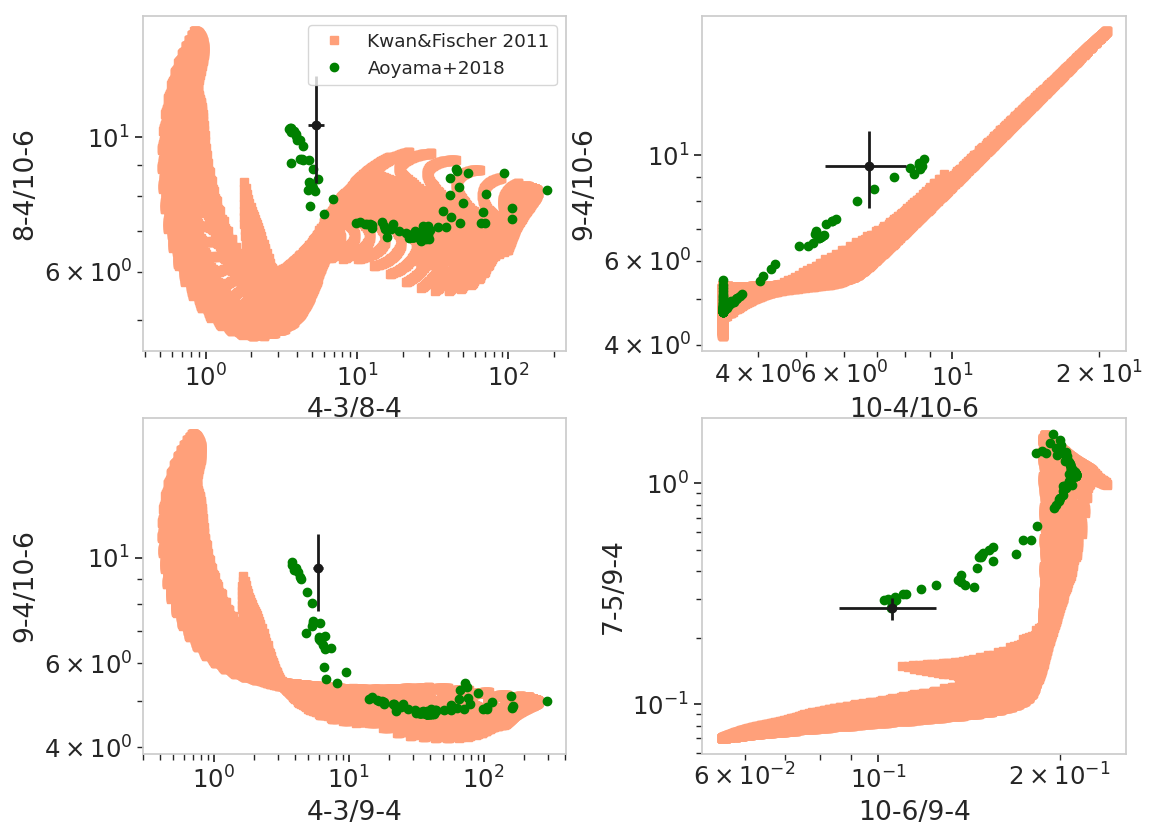}
        \caption{Ratio-ratio plot for HI transitions observed in NIRSpec. The preshock model (\citep{KF11} is shown in salmon while the postshock model \citep{Aoyama2018} is shown in green. For the 4 pairs of ratios shown here, J1605 HI emission seems to be more coherent with the postshock predictions than with the preshock ones.}
        \label{fig:ratio-ratio_NIRSpec}
\end{figure}

\begin{figure}[t]
        \centering   
        \includegraphics[width=9cm]{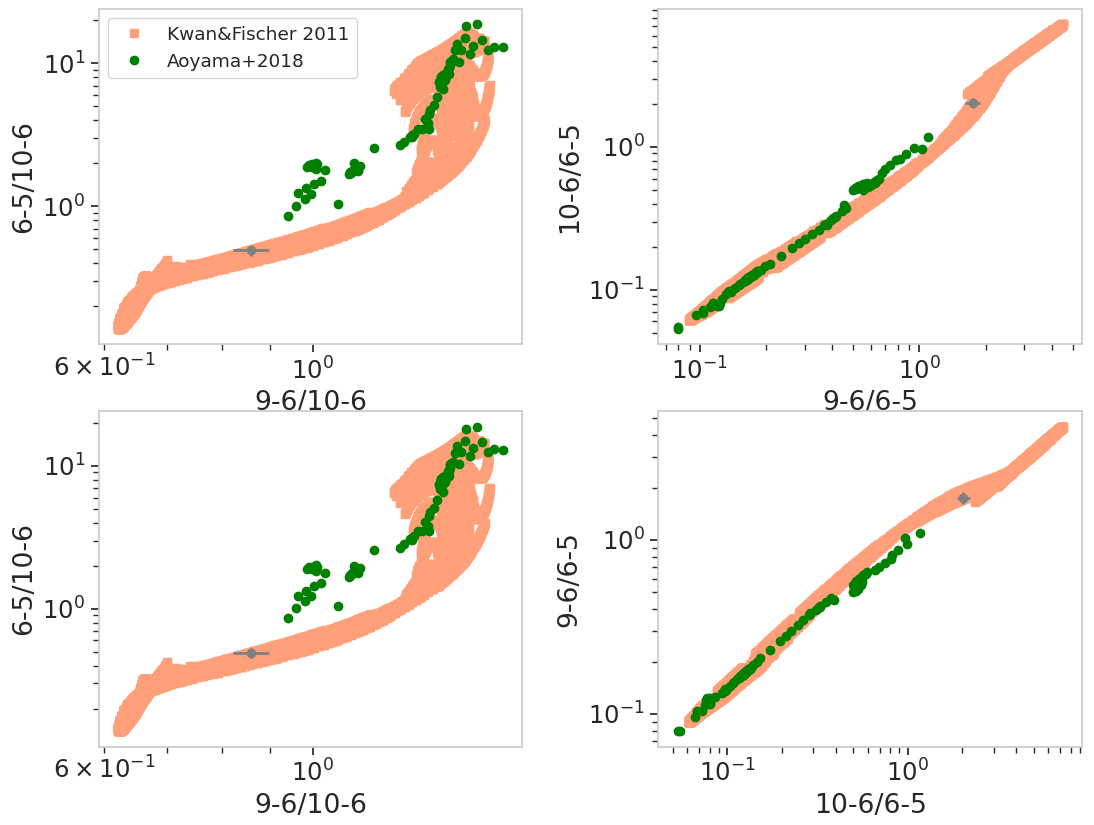}
        \caption{Same as Fig. \ref{fig:ratio-ratio_MIRI} but for the HI transitions observed in MIRI. Here, J1605's emission seems to be clearly more consistent with the emission from the preshock.}
        \label{fig:ratio-ratio_MIRI}
\end{figure}

\subsection{Fit from specific HI ratios}

In \cite{Betti2022} and \cite{Aoyama2024}, the authors used specific line ratios from both models to determine which part of the shock dominates the emission. For some key line ratios, the postshock and preshock models occupy different regions of the plot. We selected in Fig. \ref{fig:ratio-ratio_NIRSpec} (resp. Fig. \ref{fig:ratio-ratio_MIRI}) the NIRSpec (resp. MIRI) pairs of line ratios allowing us to discriminate between both models. The observed ratios at each epoch are marked with their error-bars. From this plot only, it is clear that the emission of the lines during MIRI's epoch are consistent with the preshock emission as we found earlier (Fig. \ref{fig:ratio-ratio_MIRI}). However, the NIRSpec's epoch seems to be clearly consistent with the postshock emission only, which we showed previously to be inconsistent for J1605. The key aspect here is that the models matching with our observations can be different from one ratio to the other, meaning that we cannot precisely constrain the shock parameters for all our ratios at once. The approach presented in Appendix \ref{app:postshock} prevents relying on this by investigating all the line ratios simultaneously. In general, if J1605's emission is in an intermediate stage where the emission of the shock has two notable contributions from both the postshock and the preshock simultaneously, it could produce the inconsistency we observe here. It becomes therefore crucial to derive models taking both aspects into account simultaneously and to derive grids of models to compare to observations.

\end{appendix}


\end{document}